\documentclass[aps,showpacs,twocolumn,amsmath,amssymb,eqsecnum]{revtex4}
\usepackage{graphicx}
\usepackage{color}
\usepackage{epstopdf}

\begin{document}
\title{Finite-size effects in presence of gravity: The behavior of the susceptibility \\ in $^3He$ and $^4He$ films near the liquid-vapor critical point}
\author{Daniel Dantchev$^{1,2}$\thanks{e-mail:
daniel@imbm.bas.bg},  Joseph Rudnick$^{2}$\thanks{e-mail:
jrudnick@physics.ucla.edu}, and M. Barmatz$^{3}$\thanks{e-mail:
Martin.B.Barmatz@jpl.nasa.gov}} \affiliation{ $^1$Institute of
Mechanics - BAS, Academic Georgy Bonchev St. building 4,
1113 Sofia, Bulgaria,\\
$^2$ Department of Physics and Astronomy, UCLA, Los Angeles,
California 90095-1547, USA,\\
$^3$ Jet Propulsion Laboratory, California Institute of Technology,
Pasadena, California 91109-8099, USA}
\date{\today}

\begin{abstract}
We study critical point finite-size effects on the behavior of susceptibility of a film placed in the Earth's gravitational field. The fluid-fluid and substrate-fluid interactions are characterized by van der Waals-type power law tails, and the boundary conditions are consistent with bounding surfaces that strongly prefer the liquid phase of the system. Specific predictions are made with respect to the behavior of $^3$He and $^4$He films in the vicinity of their respective liquid-gas critical points. We find that for all film thicknesses of current experimental interest the combination of van der Waals interactions and gravity leads to substantial deviations from the behavior predicted by models in which all interatomic forces are very short ranged and gravity is absent. In the case of a completely short-ranged system exact mean-field analytical expressions are derived, within the continuum approach, for the behavior of both the local and the total susceptibilities.

\end{abstract}

\maketitle

\section{Introduction}

In the case of the liquid-gas critical point, gravitational effects become important as the isothermal compressibility (i.e. the susceptibility) increases to a large value in the vicinity of that point and, indeed, diverges as that point is approached. This leads to a vertical gravity-induced density gradient that grows as the critical point is approached. Even in a relatively small experimental cell with a vertical dimension of $0.5$ mm, filled with $^3$He at its critical density, the density stratification between the cell top and bottom is approximately $6\%$  at a reduced temperature of $t = 10^{-5}$  \cite{B2000}. Only the density at the middle of the cell remains at its critical value. In order to interpret precision critical point measurements, one must develop models that incorporate the effects of gravity. A number of theoretical studies have been performed on the effect of gravity on measurements near the superfluid transition in $^4$He. Models of the $^4$He specific heat in bulk and finite-size samples near the superfluid transition have been successfully tested by high precision measurements in ground-based laboratories \cite{LC83,MG97} and in microgravity \cite{LNSSC2003,LSNGWSCIL2000}.

In a previous article \cite{DRB2007}, the authors of this paper reported the results of a theoretical calculation of the expected finite-size effects on the isothermal susceptibility near a liquid-gas critical point in a zero gravity environment. That investigation was performed for a thin film between surfaces that both strongly prefer the liquid phase. It is hoped that predictions of this study will be tested by future experimental studies in space \cite{BHLD2007}. In the present paper, we extend those calculations including the effects of gravity. The $^3$He and $^4$He liquid-gas critical points were chosen for this theoretical investigation because these systems are devoid of impurities and many bulk thermophysical properties have been measured near their liquid-gas critical point.

In this article we will discuss the behavior of the susceptibility of a film  of a non-polar fluid of, say, $^3$He or $^4$He, having a thickness $L$ in which the intrinsic interaction $J^l$ is of the van der Waals type, decaying with distance $r$ between the molecules of the fluid as $J^l \sim r^{-(d+\sigma)}$. Here $d$ is the dimensionality of the system while $\sigma>2$ is a parameter characterizing the decay of the interaction. The film is bounded by a substrate, say Au plates, that interacts with the fluid with similar van der Waals type forces, i.e. of the type $J^{l,s}\sim z^{-\sigma_s}$, where $z$ is the distance from the boundary of the system while $\sigma_s>2$ characterizes the decay of the fluid-substrate potential. For realistic fluids $d=\sigma=\sigma_s=3$.

According to finite-size scaling theory \cite{DRB2007,DSD2007,RI93}, the behavior of the susceptibility in a film of a fluid placed in an external gravitational field, governed by dispersion forces and subject to  $(+,+)$ boundary conditions, i.e., conditions that strongly favor the liquid phase of the fluid over the gas one, is
\begin{widetext}
\begin{eqnarray} \label{hypoffss}
\chi(t,\Delta\mu,L) &-& \chi_{\rm bulk}(t,\Delta\mu)-L^{-1}\left[\chi_{{\rm
surface}}^{\rm top}(t,\Delta\mu)+\chi_{{\rm surface}}^{\rm bottom}(t,\Delta\mu)\right]= \\
&& L^{\gamma/\nu}X(L/\xi_t,a_\mu (\beta\Delta\mu)L^{\Delta/\nu},a_g (\beta g) L^{1+\Delta/\nu}, h_{w,s} L^{-\omega_s}, b
L^{-\omega_b},a_\omega
L^{-\omega}), \nonumber
\end{eqnarray}
\end{widetext}
where $\chi_{\rm bulk}(t,\Delta\mu)$ is the bulk susceptibility, $\chi_{{\rm surface}}^{\rm top}(t,\Delta\mu)$ and $\chi_{{\rm surface}}^{\rm bottom}(t,\Delta\mu)$ are the surface susceptibilities---the result of the existence of two  surfaces bounding the system---$\xi_t(T)=\xi_\infty(T \rightarrow T_c^+,\Delta\mu=0)\simeq \xi_0^+|t|^{-\nu}$ is the bulk correlation length, $t=(T-T_c)/T_c$ is the reduced temperature, $T_c$ is the bulk critical temperature, $\Delta \mu=\mu-\mu_c$ is the excess chemical potential, while $\mu_c$ is the bulk critical chemical potential, $g$ is the external gravitational field, $\xi_0^+$,  $a_\mu, a_g, h_{w,s}, b$ and $a_\omega$ are nonuniversal metric factors, and
\begin{equation}\label{omegasdef}
\omega_s=\sigma_s-(d+2-\eta)/2, \qquad \omega_b=\sigma+\eta-2.
\end{equation}
The quantities $\nu$, $\Delta$, $\eta$ and $\omega$ are the universal critical exponents for the corresponding short-range system.

With respect to their bulk critical behavior, the nonpolar classical fluids belong to the so-called Ising, or $O(1)$, universality class. When $d=3$ this universality class is characterized by critical exponents \cite{JJ2002}
\begin{equation}\label{cre1}
\eta= 0.034, \gamma=1.2385, \nu=0.631,
\end{equation}
and
\begin{equation}\label{cre2}
\alpha=0.103, \beta=0.329, \theta \equiv \omega \nu =0.53.
\end{equation}
As we have already stressed, the methods and ideas to determine the behavior of the susceptibility if thin films of non-polar fluids used in the current article are general and can, in principle, be applied to any non-polar fluid; however,  but we will exemplify these methods via the study of $^3$He and $^4$He films.

A central question pertains to the extent to which the critical exponents listed above describe the experimentally observed behavior of these substances near their respective liquid-vapour critical points. In the early years of the development of critical behavior, there were attempts to experimentally measure the critical exponents of the He isotopes to be compared with theoretically predicted values;  see, e.g., \cite{BM72}.  However, it became clear that the asymptotic power-law region is very limited in ground-based measurements and accurate analyses must take into account correction-to-scaling terms as well as gravity effects; see, e.g., \cite{HM76}.   These issues are particularly important in the case of the He isotopes, which have the largest gravity effect (see, e.g.,  the discussion of Table I in \cite{BHLD2007}). Because of these problems, most recent experimental measurements, particularly those in the He isotopes, have been compared to models using the theoretical critical parameters obtained from those models; see\cite{ZB2004}.  In light of this history, we will utilize theoretically derived values of the relevant critical exponents.


From (\ref{omegasdef}) and with $\sigma=\sigma_s=3$, as in physical van der Waals interactions, one has $\omega_s=0.52$ and $\omega_b=1.03$. Since $\omega_l>\omega>\omega_s>0$, for $L$ {\it large enough}, one can expand the r.h.s. of Eq. (\ref{hypoffss}) with the result that the leading finite-size behavior of the susceptibility near the bulk critical point is given by the properties of the corresponding short-ranged system. This is true because all the dependences on the long-ranged tails of the van der Waals interaction in (\ref{hypoffss}) are  reflected via the factors $b$, $h_{w,s}$ and $a_\omega$: $b$ is proportional to the strength of the fluid-fluid interaction $J^l$ \cite{DR2001,D2001}, $h_{w,s}$ reflects the contrast between the fluid-fluid $J^l$ and the substrate-fluid $J^{l,s}$ effective interaction at $T_c$ \cite{DRB2007,DSD2007} (see below), while the field $a_w$, associated with the Wegner-type corrections to scaling, in general incorporates contributions due to the long-ranged tails of the interaction  \cite{DDG2006}. In  \cite{DRB2007,DSD2007} it is demonstrated that such an expansion is admissible only when $L$ is much larger than some critical thickness $L_{\rm crit}$, i.e., when
\begin{equation}\label{cL}
L\gg L_{\rm crit} \equiv \xi_0^+ \left(
2^{\sigma+1}|h_{w,s}|\right)^{\nu/\beta} \simeq 200 \, \xi_0^+ \,
|h_{w,s}|^{1.918}\; \mbox{\AA}.
\end{equation}
For most systems $\xi_0^+$ is of the order of 3 {\AA} and $h_{w,s}$ is of order of 1. For some systems, e.g., like $^3$He and  $^4$He bounded by Au the dimensionless constant $h_{w,s}$ can be as large as 4 \cite{DRB2007,DSD2007}. The constraint (\ref{cL}) represents the {\it ``relevance-irrelevance'' criterion} for the van der Waals forces with respect to the behavior of finite-size quantities in van der Waals thin films; when $L\gg L_{\rm crit}$ such forces can be neglected, while when $L<L_{\rm crit}$ they must be taken into account in, say, the determination of the behavior of the finite-size susceptibility. One can also formulate a criterion for the relevance of gravity. From  Eq. (\ref{hypoffss}) it is clear, that gravity is a relevant variable. If, however, $(\beta g)L^{\Delta/\nu+1} \ll 1$  the influence of gravity can be neglected and will play no essential role in the behavior of any finite-size quantity. In the opposite case its role is crucial and must be taken into account. Note that this criterion also connects the relevance of gravity to the thickness $L$ of the films; for thin films it is negligible, while in the case of sufficiently thick films it is not. In the case of $^3$He and $^4$He we will discover that films with $L=1000, 2000$ or $4000$ liquid layers are, in this sense, thin films, while a film with, say, $L=8000$ layers, can be considered thick for the purposes of assessing the influence of gravity on critical point behavior.

\begin{figure}[htb]
\includegraphics[width=\columnwidth]{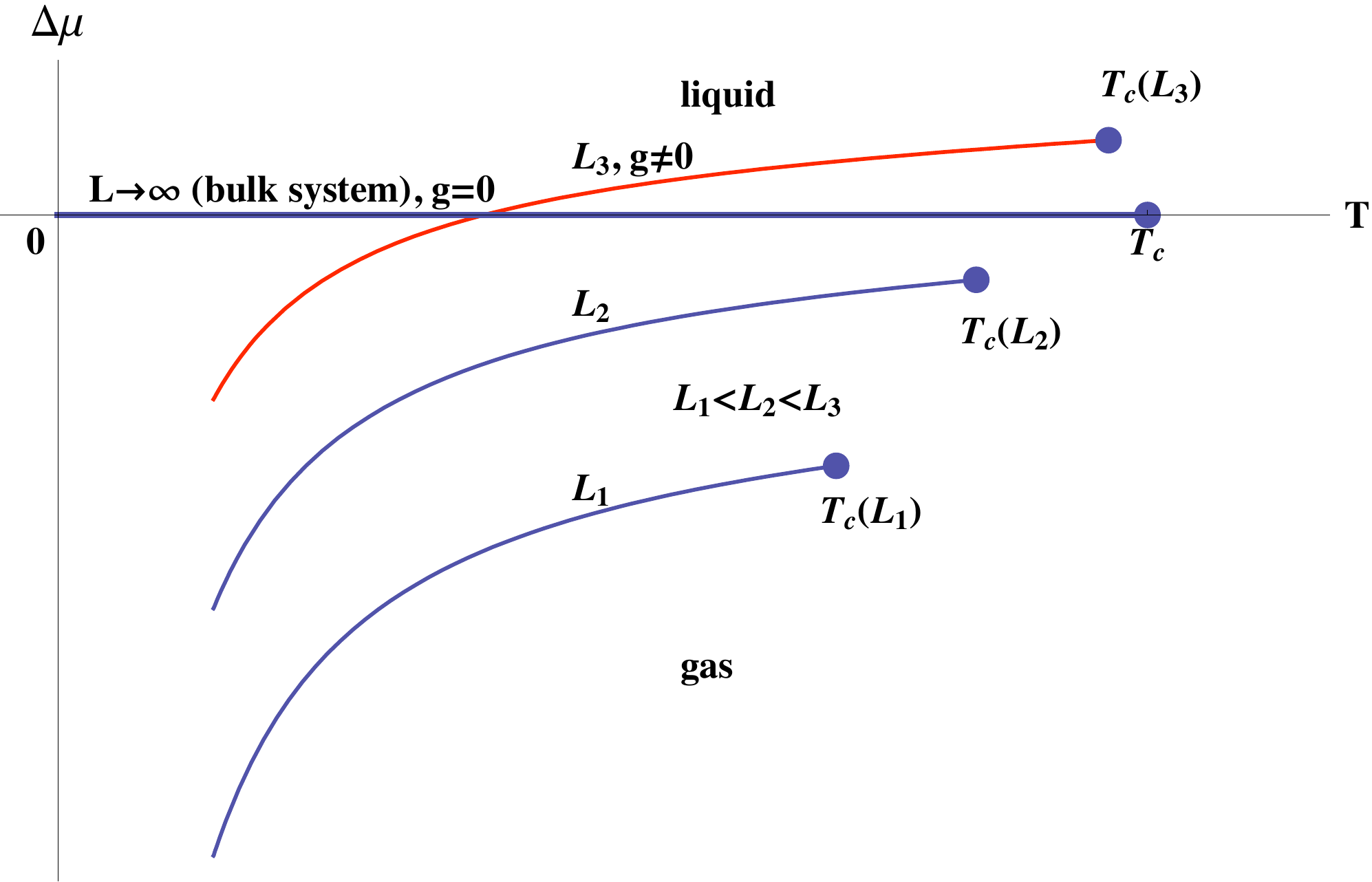}
\caption{Color online. The schematic phase diagram of a $d$-dimensional film system for various thicknesses $L$ in the presence of a gravity subject to boundary conditions that strongly favor the liquid phase at the plates bounding the fluid.
  \label{phasediagram}}
\end{figure}

In this article we discuss the finite-size behavior of the susceptibility of van der Waals fluid films bounded by two flat substrate plates situated perpendicular to the Earth's gravitational field (i.e. horizontal), both of which  strongly prefer the liquid phase of the system. The schematic phase diagram of such a system in the ($T$, $\Delta\mu$) plane is shown in Fig. \ref{phasediagram}.
The solid line $\Delta\mu=0$, $T<T_c$ represents the bulk gas-liquid phase coexistence line. The liquid-gas coexistence curves $\Delta\mu_{\rm cap}$ correspond to capillary  condensation transitions for $L=L_1$, $L=L_2$, and $L=L_3$ where $L_1<L_2<L_3$.  The case $g=0$ has been extensively studied, and the picture presented here is in accord with the results of Refs.\cite{NF82,NF83,BLM2003,BL91}. When no gravity is present within the system,  the fact that $\Delta \mu_{c,L_i}<0$ simply expresses the preferences of the identical walls (see the cases $L_1$ and $L_2$). In the presence of a gravitational field orthogonal to the bounding plates, the action of gravity pulls the molecules of the fluid away from the upper plate, leading to a more gas-like phase near that plate and a more condensed liquid-like phase region near the lower one. The larger $L$ and $g$, the stronger this effect. Thus, for $L$ large enough (see the case with $L=L_3$) the critical point of the finite system lies above the $\Delta \mu =0$ line, i.e. at $\Delta\mu_{c,L} >0$, which stabilizes the liquid phase of the fluid. Away from the critical region, the shift in the phase boundary relative to the bulk coexistence line $\Delta \mu=0$ is proportional to $L^{-1}$, while within the critical region it is proportional to
$L^{-\Delta/\nu}$ where $\Delta$ and $\nu$ are the standard bulk critical exponents. The lines of first-order phase transitions  end at $(d-1)$-dimensional critical points $T_c(L_i)$ with coordinates $(T_{c,L_i}, \Delta \mu_{c,L_i})$, $i=1,2,3$, the positions of which vary with $L$ and depend on the presence  of gravity $g$, as well as on the presence and on the strengths of the fluid-fluid and the substrate-fluid interactions.

\begin{figure}[htb]
\includegraphics[width=\columnwidth]{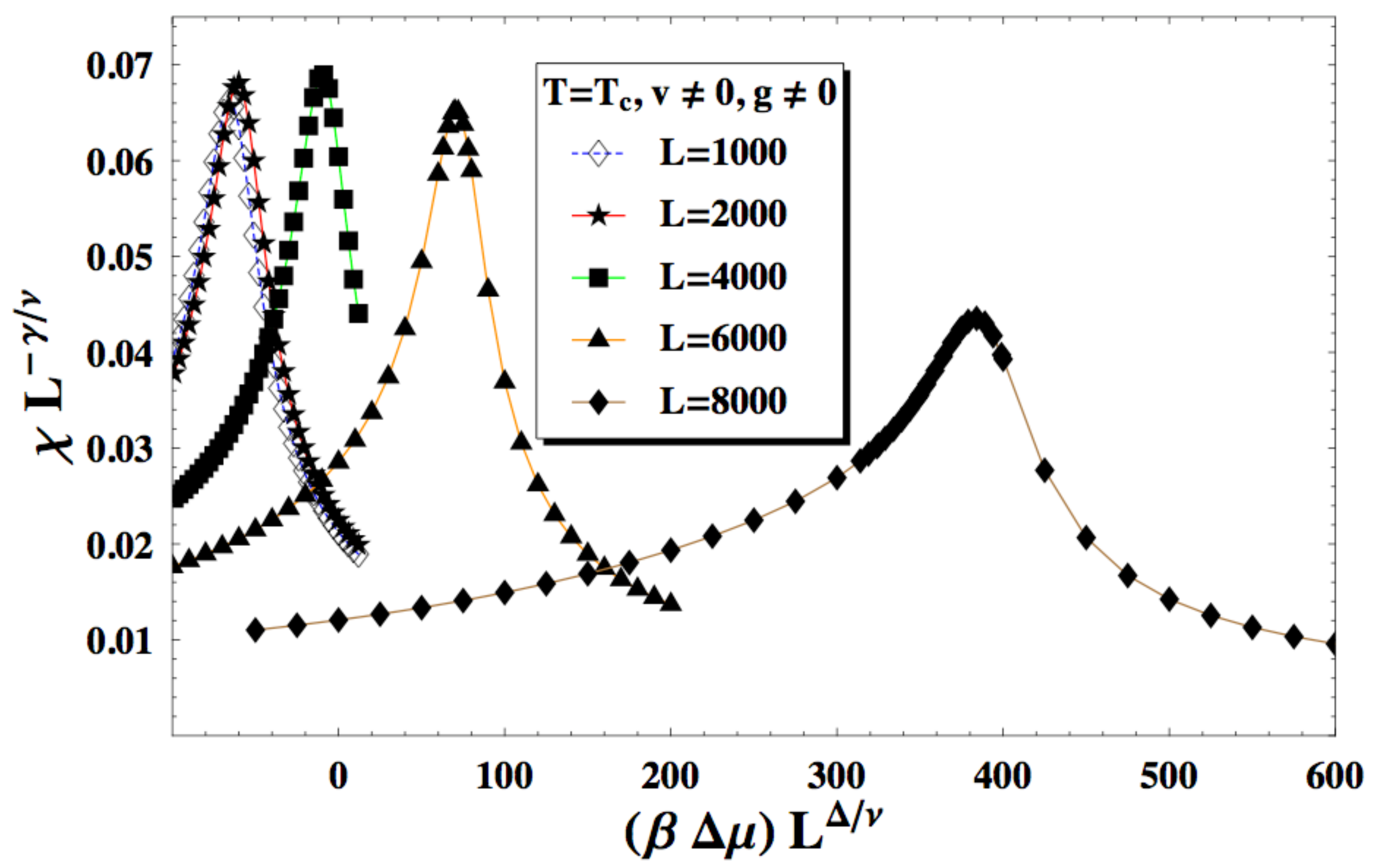}
\caption{Color online. The behavior of the finite-size susceptibility $\chi$ in thin films with thickness $L=1000, 2000, 4000, 6000$ and $8000$ layers at the bulk critical temperature $T=T_c$ of the corresponding infinite system as a function of the scaling variable $x_\mu=(\beta \Delta\mu) L^{\Delta/\nu}$. The corresponding scaling variable that governs the dependence on the gravity is proportional to $x_g\sim (\beta g) L^{\Delta/\nu+1}$ and, thus, the gravitational effects gradually set in with increase of $L$ in the behavior of the finite-size susceptibility.
  \label{chi_verus_mu_all}}
\end{figure}

\begin{figure}[htb]
\includegraphics[width=\columnwidth]{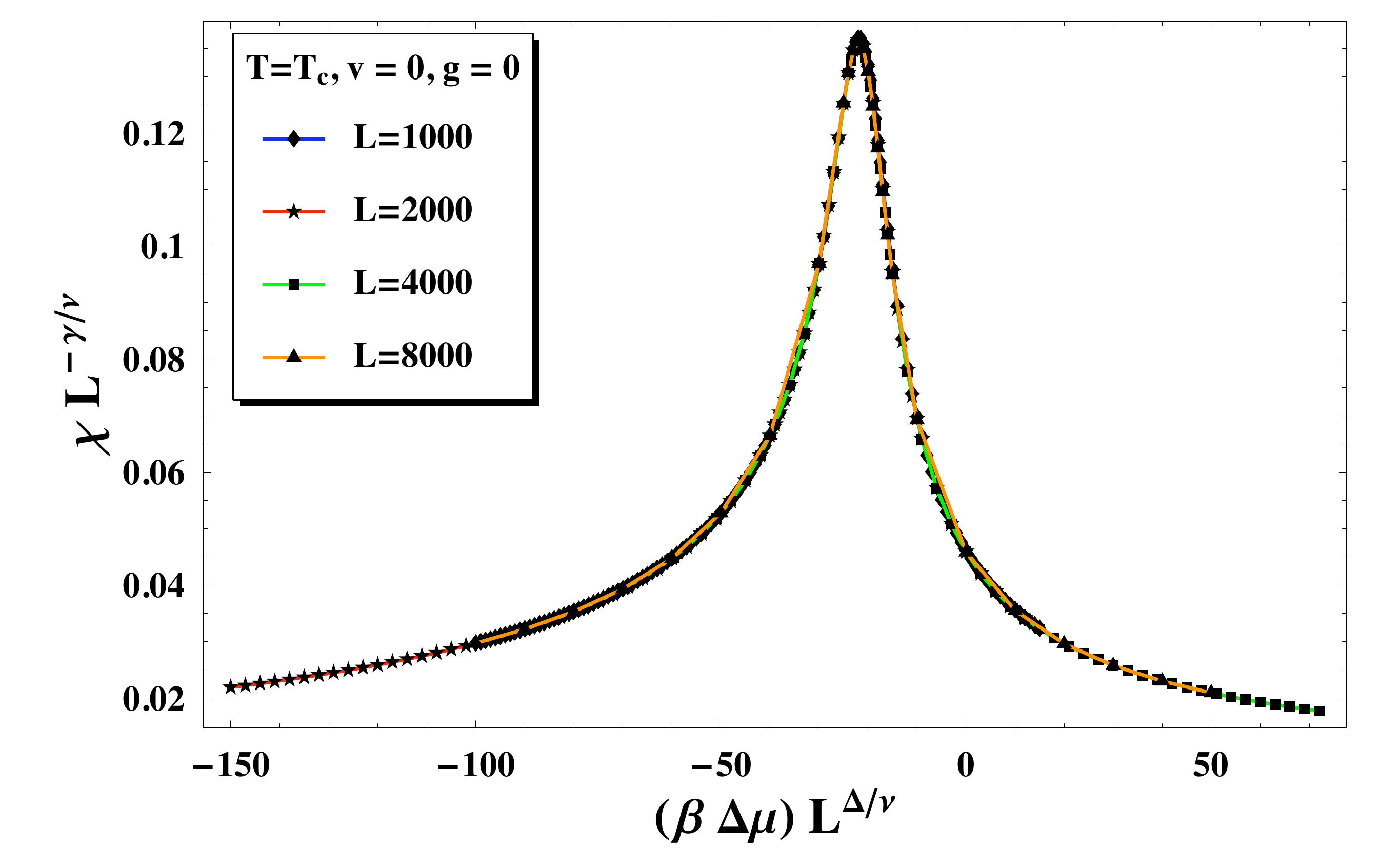}
\caption{Color online. The behavior of the finite-size susceptibility $\chi$ at $T=T_c$ in thin films with short-ranged interaction and with $L=1000, 2000, 4000$ and $8000$ layers. Note that the maximum of the susceptibility is at negative values of $x_\mu=(\beta\Delta\mu)L^{\Delta/\nu}$ Note the excellent scaling in the behavior of the curves for the different $L$'s, contrary to the behavior of the same system but with gravity and van der Waals interactions with the bounding the system plates taken into account - see the previous figure.
  \label{chi_verus_mu_sr_all}}
\end{figure}

For large $L$ these points are located close to the bulk critical point $T_c$ with
coordinates $({T_c,\Delta \mu=0})$: $T_{c,L}-T_c \sim L^{-1/\nu}$ and $\Delta \mu_{c,L}-\frac{1}{2} \beta g L\sim L^{-\Delta/\nu}$.  Since the fluctuations in systems of reduced size are stronger, one typically has $T_{c,L_i}<T_c$.  In part, the structure of this phase diagram is reflected in Fig. \ref{chi_verus_mu_all}, where the behavior of the finite-size susceptibility in thin films with thickness $L=1000, 2000, 4000, 6000$ and $8000$ layers is shown as a function of the scaling variable $x_\mu=(\beta \Delta\mu) L^{\Delta/\nu}$ at the bulk critical temperature $T=T_c$ of the corresponding infinite system (with $g=0$). Both the presence of the van der Waals interaction between the fluid particles and between the substrate and the fluid, as well as the gravitational field of the Earth are taken into account. One observes a clear lack of data collapse.  For relatively thin films---with $L=1000$, $2000$ and $4000$---this is due to the role of the van der Waals interaction, while for relatively thick films with, say, $L=8000$ the absence of data collapse is due to the presence of gravity. The $L=6000$ case illustrates the intermediate situation when the influence of van der Waals interactions fades away and the gravity steps in as a factor mainly responsible for the lack of data collapse.  When neither gravity nor van der Waals type interactions are present, a perfect data collapse can be achieved, see Fig.~\ref{chi_verus_mu_sr_all},  where the data are plotted in the same way as in the current figure. Note also that both the van der Waals interactions and the presence of gravity reduces the magnitude of the susceptibility. This is due to the ordering effect of the van der Waals interactions and gravitational field. In the cases $L=1000, 2000$ and $4000$, the van der Waals interactions constitute the important influence leading to the effect, and
this reduction is approximately given by a factor of 2. On the other hand,
when $L=8000$ the reduction in the maximum value of the susceptibility is by a
factor of 3 times and is primarely due to the influence of gravity.
We also note that the maximum of the susceptibility as a function of $\Delta\mu$
also changes its location; while for $L=1000, 2000$ and $4000$ the maximum
occurs when  $\Delta \mu <0$ with $x_\mu^{\rm max} \equiv (\beta\Delta\mu)L^{\Delta/\nu}=O(1)$,
for $L=6000$ and $8000$ the maximum is at $\Delta \mu >0$ with $x_\mu^{\rm max}\sim \beta_c g L/2 \gg 1$.
One of the aims of the current article is to explain how the above curves have been obtained and to elucidate why, as we believe,  these curves ought to resemble the ones obtained in experiments with real liquid-gas systems. In order to facilitate contact with experiment we will, in addition to presenting the behavior of $\chi$ in the $(T,\Delta\mu)$ plane, derive the corresponding dependence of $\chi$ in the $(T,\Delta\rho)$ plane, where $\Delta\rho=\rho-\rho_c$, with $\rho_c$ the critical density of the fluid.

The structure of the article is as follows. First, in Section \ref{model} we present a precise formulation of the model of interest. The corresponding simplification of the model in the case of a film geometry and the analytical expressions needed for its numerical treatment are presented in Section \ref{film}.
The results for the behavior of the finite-size susceptibility at the critical point as a function of $\mu$, $L$ and $g$ are presented in Section \ref{at_criticality}, while the corresponding results for $T>T_c$ and $T<T_c$ are given in Section \ref{off_criticality}. The article closes with a discussion and concluding remarks.

\section{The model}
\label{model}

Following Refs. \cite{DRB2007} and \cite{DSD2007} we consider a lattice-gas model of a fluid confined between two parallel flat plates at a distance $L$ with a grand canonical functional $\Omega[\rho({\bf r})]$ given by
\begin{eqnarray}\label{freeenergyfunctionalstarting}
\Omega[\rho({\bf r})]&=& k_B T \sum_{{\bf r}\in {\cal L}} \Big\{
\rho({\bf r}) \ln\left[\rho({\bf r})\right]\nonumber \\ && +
[1-\rho({\bf r})] \ln\left[1-\rho({\bf r})\right]\Big\}\nonumber \\
&& +\frac{1}{2}\sum_{{\bf r}, {\bf r}'\in {\cal L}}\rho({\bf
r})w^{(l)}({\bf r}-{\bf r}')\rho({\bf r}') \nonumber \\ &&
 +\sum_{{\bf r} \in {\cal L}}[V^{\rm (l,s)}(z)+ g z -\mu] \rho({\bf
r}).
\end{eqnarray}
This expression is to be minimized with respect to the local number density $\rho({\bf r})$. The functional (\ref{freeenergyfunctionalstarting}) is the simplest model that captures the basic features of systems with both van der Waals interactions and gravity taken into account. It can be viewed as a modification of the model utilized by Fisher and Nakanishi in their mean-field investigation of short-range systems \cite{NF82,NF83} and in the absence  of gravity.

In Eq. (\ref{freeenergyfunctionalstarting}) $w^{(l)}({\bf r}-{\bf r}')=-4J^{(l)}({\bf r}-{\bf r}')$ is the non-local coupling between the constituents of the fluid, while ${\cal L}$ is a simple cubic lattice in the region
occupied by the fluid.  We consider a fluid system with geometry $\infty^{d-1}\times [0,L]$ where the region $0 \le z \le L$ is occupied by fluid. Here and in the remainder of this paper, all length scales are taken in units of the lattice constant $a$, which is of the order of a molecular diameter---which means that length is expressed as a dimensionless quantity---so that the particle density
$\rho({\bf r})$ is dimensionless and varies within the range $[0,1]$. In Eq.\,(\ref{freeenergyfunctionalstarting}), the terms in curly brackets correspond to the entropic contributions, while the other terms are directly related to the interactions present in the system. The term proportional to $g$ reflects the presence of gravity. It is assumed that the gravitational field is along the $z$ direction,
i.e. is perpendicular to the  planes bounding the fluid. The external potential $V^{\rm (l,s)}(z)$ reflects the interaction between the molecules of the fluid and of the constituents of the two substrates bounding it.  For an individual wall $V^{\rm (l,s)}(z\rightarrow\infty)\sim z^{-\sigma}$ with $\sigma=3$ for a
genuine van der Waals interaction. In the current treatment we will assume that the bounding substrates are identical on the both sides of the film and that they strongly prefer the liquid phase of the fluid. In Eq.\,(\ref{freeenergyfunctionalstarting}), $\mu$ is the chemical potential.

The variation of Eq. (\ref{freeenergyfunctionalstarting})  with respect to $\rho({\bf r})$ leads to the equation of state for the equilibrium density $\rho^*({\bf r})$
\begin{eqnarray}
\label{eqofstate}
  2\rho^*({\bf r})-1  &=&\tanh \Bigg[-\frac{\beta}{2}\sum_{{\bf r}'}
    w^{(l)}({\bf r}-{\bf r}')\rho^*({\bf r}') \nonumber \\ &&
    +\frac{\beta}{2}\left(\mu -V^{\rm (l,s)}(z)-g z\right)
    \Bigg].
\end{eqnarray}
The advantage of this equation is that it lends itself to numerical solution by iterative procedures. For a particular geometry and surface potential $V^{\rm (l,s)}(z)$ the solution determines the equilibrium order-parameter profile $\rho^*({\bf r})$ in the system. Inserting this profile into Eq.\,(\ref{freeenergyfunctionalstarting}) one obtains the system's grand potential. To avoid the double
sum in Eq.\,(\ref{freeenergyfunctionalstarting}), which is inconvenient in a numerical treatment, we make use of the relationship below, which is easily derived from Eq.
(\ref{eqofstate})
\begin{eqnarray}\label{asum}
\lefteqn{\frac{1}{2}\sum_{{\bf r}, {\bf r}'\in {\cal
L}}w({\bf r}-{\bf r}')\rho^*({\bf r})\rho^*({\bf r}')} \\
&=&\frac{1}{2}\sum_{{\bf r}\in {\cal L}}\left[ \mu -
V(z)-g z\right] \rho^*({\bf r}) \nonumber \\
&&-k_B T \sum_{{\bf r}\in {\cal L}}\rho^*({\bf r})\  {\rm
arctanh[2\rho^*({\bf r})-1]},\nonumber
\end{eqnarray}
which, when inserted in Eq.\,(\ref{freeenergyfunctionalstarting}), yields
\begin{eqnarray}\label{freeenergyfinal}
\!\!\!\!\!\!\! \!\!\!&& \Omega[\rho^*({\bf r})]=\sum_{{\bf r}\in
{\cal L}}\Bigg[ k_BT\left\{ \rho^*({\bf r}) \ln\left[\rho^*({\bf
r})\right] \right. \nonumber
\\ \!\!\!\!\!\!\! \!\!\! &&  \left.+ [1-\rho^*({\bf r})] \ln\left[1-\rho^*({\bf r})\right]-
\rho^*({\bf r})\ {\rm arctanh[2\rho^*({\bf r})-1]} \right\}
\nonumber \\ \!\!\!\!\!\!\! \!\!\! && -\frac{1}{2}
\left[\mu-V(z)-g z\right] \rho^*({\bf r})\Bigg].\!\!\!\!\!
\!\!\!\!\!\!\!\!\!\!\!\!
\end{eqnarray}
Note that in (\ref{freeenergyfinal}) $\rho^*({\bf r})$ is no longer a free functional variable,  but is the solution of Eq.\,(\ref{eqofstate}).

Denoting $\phi^*({\bf r})=2 \rho^*({\bf r}) -1$ and $\Delta \mu=\mu-\mu_c$, where $\mu_{c}=\frac{1}{2} \sum_{{\bf r}'} w({\bf r}-{\bf r}')$, the equation of state (\ref{eqofstate}) can be
rewritten in the standard form
\begin{eqnarray}\label{eqstatestandard}
\phi^*({\bf r}) &=& \tanh \Bigg[ \beta \sum_{{\bf r}'} J({\bf r}, {\bf
r}') \phi^*({\bf r}') \nonumber \\
&& +\frac{\beta}{2}\Big(\Delta \mu - \Delta
V(z)- g z\Big)\Bigg],
\end{eqnarray}
where $J({\bf r}-{\bf r}')=-w({\bf r}-{\bf r}')/4 $. The bulk properties of the model are well known (see, e.g.,\cite{D96,B82} and references therein). We recall that the order parameter $\phi^*$ of the system has a critical value $\phi^*=0$ which corresponds to $\rho_c=1/2$ so that $\phi^*=2(\rho^*-\rho_c)$.
The bulk critical point of the model is given by $(\beta=\beta_c=[\sum_{\bf r}J({\bf r})]^{-1}, \mu=\mu_c=-2\sum_{\bf r}J({\bf r}))$ with the sum running over the whole lattice. Within the mean-field approximation the critical exponents for the order parameter and the compressibility are $\beta=1/2$ and $\gamma=1$, respectively. As has been shown in \cite{DSD2007,DRB2007}, the surface potential $\Delta V(z)$  is
\begin{equation}
\label{deltav} \Delta V(z)= \delta v_s
\left[(z+1)^{-\sigma}+(L+1-z)^{-\sigma}\right],
\end{equation}
where $1\le z\le L-1$, and where contributions of the order of $z^{-\sigma-1}$, $z^{-\sigma-2}$, etc. have been neglected, the quantity
\begin{equation}\label{sdef}
\delta v_s=
-4\pi^{(d-1)/2}\frac{\Gamma\left(\frac{1+\sigma}{2}\right)}
{\sigma\Gamma\left(\frac{d+\sigma}{2}\right)}(\rho_s J^{l,s}-\rho_c
J^l)
\end{equation}
is a  ($T$- and $\mu$-independent) constant, the quantity
\begin{equation}
\label{Jfluid} J({\bf r})\equiv  J^l/(1+|{\bf r}|^{d+\sigma}),
\end{equation}
is a proper lattice version of $-w({\bf r})/4$ as the interaction
energy between the fluid particles, and
\begin{equation}
\label{Jsubstrate} J^{l,s}({\bf r})\equiv  J^{l,s}/|{\bf r}|^{d+\sigma}
\end{equation}
is the interaction between a fluid particle and a substrate particle. Here $\rho_s$ is the number density of the substrate particles in units of $a^{-d}$.  Note that the effective potential $\delta v_s$ is the result of the difference between the relative strength  of the substrate-fluid interaction for a substrate with density $\rho_s$ and that of the fluid-fluid interaction  for a fluid with a density $\rho_c$.
In Eq. (\ref{deltav}),
the restriction $z\geq 1$ holds because we consider the layers closest to the
substrate to be completely occupied by the liquid phase of the fluid,
which implies that we consider the strong adsorption limit, i.e., $\rho(0)=\rho(L)=1$;
therefore the actual values of $\Delta V(0)=\Delta V(L)$ will play no role.

In terms of the quantity $\phi$ the functional (\ref{freeenergyfunctionalstarting}) becomes
\begin{eqnarray}\label{gcpomega}
\Omega[\phi({\bf r})]&=& k_B T \sum_{{\bf r}\in {\cal L}} \Bigg\{
\frac{1+\phi({\bf r})}{2} \ln\left[\frac{1+\phi({\bf
r})}{2}\right]\nonumber \\ && + \frac{1-\phi({\bf r})}{2}
\ln\left[\frac{1-\phi({\bf r})}{2}\right]\Bigg\} \nonumber \\ &&
 -\frac{1}{2}\sum_{{\bf r} \in {\cal L}}[\Delta \mu- \Delta V(z)- g z] \phi({\bf
r}) \\
& & -\frac{1}{2}\sum_{{\bf r}, {\bf r}'\in {\cal L}}J({\bf r},{\bf
r}')\phi({\bf r})\phi({\bf r}')+\Omega_{\rm reg},\nonumber
\end{eqnarray}
where
\begin{equation}\label{regomega}
\Omega_{\rm reg}=-\frac{1}{2}\sum_{{\bf r} \in {\cal L}}\left[\Delta
\mu-\Delta V(z)-g z-\sum_{{\bf r}'\in {\cal L}}J({\bf r}, {\bf
r}')\right]
\end{equation}
does not depend on $\phi$ and therefore is a regular background term.  An expression similar to the one in Eq. (\ref{freeenergyfinal}), which avoids    the double sum and thus is   more convenient for numerical procedures, can also be  obtained. With the identifications
$\phi({\bf r})\leftrightarrow m({\bf
 r})$ and
\begin{equation}\label{Hdef}
H(z)=\frac{1}{2}[\Delta \mu -\Delta V(z)-g z]
\end{equation}
one can rewrite the above expression for  $\Omega[\phi({\bf r})]$ as a functional of the effective magnetic density $m({\bf r})$:
 \begin{equation}\label{fdef}
    F[m({\bf r})] \equiv (\Omega-\Omega_{\rm reg}),
 \end{equation}
 which defines the free energy of a magnetic system at temperature $T$ and in the presence of an external local and spatially varying magnetic field $H(z)$. Explicitly, one has
 \begin{widetext}
\begin{equation}\label{freeenergyfunctionalstartingm}
\beta F[m({\bf r})]=\sum_{{\bf r}}\left\{\frac{1+m({\bf r})}{2}
\ln\left[\frac{1+m({\bf r})}{2}\right]+\frac{1-m({\bf r})}{2}
\ln\left[\frac{1-m({\bf r})}{2}\right]\right\}
 -\sum_{{\bf r}}h(z) m({\bf
r})-\frac{1}{2}\sum_{{\bf r}, {\bf r}'}K({\bf r},{\bf r}')m({\bf
r})m({\bf r}'),
\end{equation}
\end{widetext}
where  $K({\bf r},{\bf r}')=\beta J^l({\bf r},{\bf r}')$ is the non-local coupling between magnetic degrees of freedom, $h(z)=\beta H(z)$ is an external magnetic field and the magnetization $m({\bf r})$ is to be treated as a variational parameter.

In the remainder of this article we shall make use of this connection between the fluid and magnetic systems in order to exploit existing theoretical results for both of them.

\section{Finite-size behavior of the model in a film geometry}
\label{film}

We will be interested in a system with a {\it film} geometry. Because of the symmetry of the system one has $\phi({\bf r})\equiv \phi({\bf r}_\|,z)=\phi(z)$, where ${\bf r}=\{{\bf r}_\parallel,z\}$, i.e., the order parameter profile $\{\phi(z),0 \le z\le L\}$, with $\phi(0)=\phi(L)=1$, depends only on the coordinate perpendicular to the plates bounding the van der Waals system. In this case Eq. (\ref{eqstatestandard})
becomes
\begin{equation}\label{mpfilm}
\phi^*(z)=\tanh\left[\beta \sum_{z'=0}^L \hat{{\cal J}}(z-z')\phi^*(z')+h(z)\right],
\end{equation}
where
\begin{equation}\label{Gdef}
\hat{{\cal J}}(z)\equiv \sum_{{\bf r}_\|'}J({\bf r}_\|-{\bf
r}_\|',z)=\sum_{{\bf r}_\|}J({\bf r}_\|,z).
\end{equation}
In \cite{DRB2007} it has been shown that the function $\hat{{\cal J}}(z)$
can be written in the form
\begin{eqnarray}\label{J}
\hat{{\cal J}}(z) &=&J^l \left[c_{d-1}\delta(z)+c_{d-1}^{\rm
nn}\left[\delta(z-1)+\delta(z+1)\right] \right. \nonumber \\
&&\left. +{\cal G}_d(z) \theta(z-2)\right],
\end{eqnarray}
where $\delta(z)$ is the discrete delta function, while $\theta(z)$ is the Heaviside function. Explicitly, for $d=\sigma=3$  one has \cite{DRB2007}
\begin{equation}\label{c2t}
c_2=\sum_{{\bf n}\in {\mathbb Z}^2}\frac{1}{1+|{\bf n}|^6}\simeq
3.602,
\end{equation}
\begin{widetext}
\begin{eqnarray}\label{c2nnt}
c_2^{nn}&=&-\frac{8}{3} \pi
\left[(-1)^{1/3}K_0(\sqrt{2-2i\sqrt{3}\pi})-(-1)^{2/3}K_0(\sqrt{2+2i\sqrt{3}\pi})\right]
+ \frac{\pi}{3}\left(\frac{\pi}{\sqrt{3}}-\ln 2\right) \approx
1.183,
\end{eqnarray}
and
\begin{equation}
\label{gt} {\cal G}_3(x)=\frac{\pi}{3}\left[\sqrt{3}\ \arctan{
\frac{\sqrt{3}}{2x^2-1}}-\ln\left(1+\frac{1}{x^2}\right)+
\frac{1}{2}\ln\left(1-\frac{1}{x^2}+\frac{1}{x^4}\right)\right],
\end{equation}
\end{widetext}
where $K_0(x)$ is a modified Bessel function of the second kind.
Following \cite{DRB2007} and taking into account the presence of gravity in the system we study, the layer magnetic field $h(z)$ is
\begin{eqnarray}\label{hz}
h(z)&=&\frac{1}{2}\beta\Delta\mu -\frac{1}{2}\beta g z \\
&& + \frac{h_{w,s}}{(z+1)^{\sigma_s}}+\frac{h_{w,s}}{(L+1-z)^{\sigma_s}},
\ \ 1\le z \le L-1, \nonumber
\end{eqnarray}
where
\begin{equation}\label{hzdef}
h_{w,s}=-\frac{1}{2}\beta \delta v_s
\end{equation}
reflects the relative strength of the fluid-wall and fluid-fluid interactions, respectively.   The above expression takes into account the fact that the substrate occupies the region $\mathbb{R}^{d-1}\times
[L+1,\infty] \; \cup \; \mathbb{R}^{d-1}\times [-(L+1),-\infty]$. For $^3$He and $^4$He bounded by Au surfaces in \cite{DRB2007} it has been shown that $h_{\rm ws}=4$.  Note that (\ref{hz}) is derived for a system with $\sigma=\sigma_s$. According to finite-size scaling theory the finite-size effects due to the surface field $h_{w,s}$ are controlled by $h_{w,s} L^{(d+2-\eta)/2-\sigma_s}$. For $d=\sigma=\sigma_s=3$, an Ising-like system, this leads to $h_{w,s}/\sqrt{L}$, where the value of $\eta=0.034$ has been neglected, i.e., $\eta=0$ was used. Let us recall that within a mean-field treatment with respect to the critical behavior,  the effective spatial dimension is $d=4$ irrespective of the actual spatial dimension of the model under consideration. In order to have within the current mean-field model the same order of the finite-size effects due to $h_{w,s}$ as in real systems we take in our model calculations  $\sigma_s=3.5$.  This value will be used in the remainder of the article whenever the substrate-fluid interaction is taken into account.

With respect to the behavior of the total susceptibility $\chi$ of the system per unit particle it was shown in \cite{DRB2007} that
\begin{equation}\chi
=\frac{1}{L+1}\sum_{z,z^*}\left(\bf{R}^{-1}\right)_{z,z^{*}},
\end{equation}
where $\bf{R}^{-1}$ is the inverse matrix of the matrix $\bf{R}$
with elements
\begin{equation}\label{R_matrix_def}
    R_{z,z'}=\frac{\delta_{z,z'}}{1-\phi^2(z')}-\beta \hat{{\cal
J}}(z-z'),
\end{equation}
while the ``local'' susceptibility, which reflects  the response of the system from a given layer is
\begin{equation}\label{chi_sol}
\chi_l(z)=\sum_{z^*}\left(\bf{R}^{-1}\right)_{z,z^{*}},
\end{equation}
with
\begin{eqnarray}\label{chideflocal}
 \chi_l(z) & \equiv & \sum_{z^*} G(z,z^*) \\ & = & \sum_{{\bf
r}^*}\langle S({\bf 0},z)S({\bf r}_\|^*,z^*)\rangle-\langle S({\bf
0},z)\rangle \langle S({\bf r}_\|^*,z^*)\rangle. \nonumber
\end{eqnarray}
Obviously, $\chi \equiv \sum_z\chi_l(z)/(L+1)$.

For a fluid confined to a film geometry
the natural quantity to consider  is the  excess grand potential normalized per unit
area $A$:  $\Delta \omega\equiv \lim_{A\to\infty} (\Omega-\Omega_{\rm reg})/A$.
From Eq. (\ref{gcpomega}) and (\ref{mpfilm}) one obtains
\begin{eqnarray}\label{gcpomegafilm}
\Delta \omega[\phi(z)]&=& k_B T \sum_{z=0}^L \Bigg\{
\frac{1+\phi(z)}{2} \ln\left[\frac{1+\phi(z)}{2}\right]\nonumber \\ && + \frac{1-\phi(z)}{2}
\ln\left[\frac{1-\phi(z)}{2}\right]\Bigg\} \nonumber \\ &&
 -\frac{1}{2}\sum_{z=0}^L [\Delta \mu- \Delta V(z)- g z] \phi(z) \\
& & -\frac{1}{2}\sum_{z=0}^L\sum_{z'=0}^L {\hat{\cal J}}(z,z')\phi(z)\phi(z').\nonumber
\end{eqnarray}
The order parameter profile of the system is the one that provides the minimum of the above
functional. For such a profile $\{\phi^*(z),0\le z \le L\}$, which is a solution of
Eq. (\ref{mpfilm}), the above expression can be simplified to
\begin{eqnarray}\label{gcpomegafilmfinal}
\beta \Delta \omega &=& \sum_{z=0}^L \Bigg\{
\frac{1+\phi^*(z)}{2} \ln\left[\frac{1+\phi^*(z)}{2}\right]\nonumber \\ && + \frac{1-\phi^*(z)}{2}
\ln\left[\frac{1-\phi^*(z)}{2}\right] \\ &&
 -\frac{1}{2}  h(z) \phi^*(z) -\frac{1}{2}\,  \phi^*(z) \; {\rm arctanh}
 \left[\phi^*(z)\right]\Bigg\},\nonumber
\end{eqnarray}
which is much more convenient for numerical evaluation since it does not involve the double
summation present in (\ref{gcpomegafilm}).

Eqs. (\ref{mpfilm}), (\ref{Gdef})-(\ref{R_matrix_def}), and (\ref{gcpomegafilmfinal})
provide the basis for our numerical treatment of the finite-size behavior of the
susceptibility of a system in which both the van der Waals interaction between
the molecules of the fluid and between the fluid and the constituents of the substrate,
as well as the Earth gravity are taken into account. The procedure is as follows.
First, we  determine the order parameter profile  $\{\phi^*(z),0 \le z \le L\}$
by solving iteratively, using  the Newton-Kantorovich method, Eq. (\ref{mpfilm}). However,
the solution of this equation depends, for a given range of parameters $T$ and $\Delta \mu$,
on the choice of the initial state of the order parameter profile.
The two basic initial states of the profile are {\it i)}
a liquid-like state in which all the sites of the lattice are occupied by a particle,
i.e. the state $\{\phi(z)=1, 0 \le z \le L\}$, and {\it ii)} a
gas-like state $\{\phi(0)=1, \phi(L)=1, \phi(z)=0, 1 \le z \le L-1\}$.
Thus one needs to calculate the profile starting from both of the two initial states.
If the  two final states coincide they provide the unique minimum of the
functional $(\ref{gcpomegafilm})$. If they differ one has to check which one
provides the absolute minimum of the grad canonical potential.  The simplest way to
clarify that question is to calculate $\beta \Delta \omega$ via Eq. (\ref{gcpomegafilmfinal}).

The standard Ginzburg-Landau equation follows from Eq. (\ref{mpfilm}), for small $\phi$, after taking into
account that ${\rm arctanh}(\phi)\simeq \phi + \phi^3/3 +O(\phi^5)$. A
continuum version of the equation follows from the replacement
$\phi(z-1)+\phi(z+1)\rightarrow 2\phi(z)+\phi''[z]$. Obviously such a
continuum version can also be constructed for the long-range
system by adding the terms contributed by the function
${\cal G}(x)$, which is, in this case, a continuous
function. Note that the function ${\cal G}(x)$ is well defined everywhere
for $x\ge 0$ and not only for $x\ge 1$ as we actually need it in
the lattice formulation of the theory. Thus, in the continuum
formulation of the theory the integration can be extended over the
region $z\in [0,L]$. This does not change the long-range behavior
of the magnetization profiles. Thus, in the continuum case
the equation for the order parameter profile reads
\begin{widetext}
\begin{equation}\label{eqcont}
\phi^*[z]+\frac{1}{3}(\phi^{*}[z])^3=h[z]+K\left\{c_2 \phi^*(z) + c_2^{nn}
\left[2 \phi^*(z)+\frac{d^2 \phi^*(z)}{dz^2}\right]+\int_{0}^{L}{\cal
G}(|z-z'|^2)\phi^*(z')dz' \right\},
\end{equation}
\end{widetext}
where $K=\beta J^l$ and $h(z)$, for $0< z < L$, is
\begin{equation}\label{hzc}
h(z)=\frac{1}{2}\beta\Delta\mu -\frac{1}{2}\beta g z+ h_{w,s}\left[{z^{-\sigma_s}}+(L-z)^{-\sigma_s}\right].
\end{equation}
In continuum theory one defines the $(+,+)$ boundary conditions via  $\phi^*(0)=\phi^*(L)=\infty$.

\subsection*{The model with purely short-range interactions for $\Delta\mu=0$ and $g=0$}
\label{psrs}

In the case in which all the interactions in the system are short-ranged and in the absence of gravity the equation (\ref{eqcont}) for the order parameter profile in a continuum system can be written in the standard form
\begin{equation} \label{eqcontsrmt}
-\frac{d^2 \phi}{dz^2}+\hat{a} \phi + u \phi^3=\hat{h},
\end{equation}
where
\begin{equation}\label{srconstmt}
\hat{a}\equiv \frac{1}{c_2^{nn}K}\left(1-\frac{K}{K_c}\right),
\quad u\equiv\frac{1}{3 c_2^{nn}K}, \quad \mbox{and} \quad \hat{h}\equiv\frac{h}{c_2^{nn}K},
\end{equation}
with
\begin{equation}\label{Kc}
    K_c=1/\left(c_2+2c_2^{\rm nn}\right).
\end{equation}
The equations for the layer response function $\chi_l(z)$, where
\begin{equation}\label{localsusdefmt}
\chi_l(z)\equiv \frac{\partial \phi(z)}{\partial \hat{h}}{{\Bigg |}_{\hat{h}=0}}
\end{equation}
then reads
\begin{equation} \label{eqcontsrchimt}
-\frac{d^2 \chi_l}{dz^2}+(\hat{a}+ 3u \phi^2) \chi_l =1.
\end{equation}
Because conditions are identical at both bounding surfaces of the
system, the solutions of the above equations have to satisfy  $\phi'(L/2)=0$ and
$\chi'(L/2)=0$. Under the $(+,+)$ boundary conditions envisaged here (strong adsorption) one has, in addition, $\phi(0)=\phi(L)=+\infty$, and $\chi(0)=\chi(L)=0$.

When $\hat{h}=0$ the magnetization profile  is known exactly
\cite{M97}:
\begin{enumerate}

\item[a)] when $x_t\equiv\hat{a}L^2\ge -\pi^2$
\begin{eqnarray}\label{mz}
\phi(z)&=&L^{-1} \sqrt{\frac{2}{u}}  \left\{2 K(k)\frac{{\rm dn}[2 K(k)\zeta;k]}{{\rm sn}[2 K(k)\zeta;k]}\right\} \nonumber \\
&=&z^{-1} \sqrt{\frac{2}{u}}  \left\{2 K(k)\zeta \frac{{\rm dn}[2 K(k)\zeta;k]}{{\rm sn}[2 K(k)\zeta;k]}\right\},
\end{eqnarray}
where $k^2 \ge 0$ is to be determined from
\begin{equation}\label{tk}
    x_t=[2K(k)]^2(2k^2-1),
\end{equation}
and $\zeta=z/L$, i.e., $0\le\zeta\le 1$.

\item[b)] when $x_t \le -\pi^2$
\begin{eqnarray}\label{mzd}
\phi(z) &=& L^{-1}\sqrt{\frac{2}{u}}\left\{\frac{2 K(\bar{k})}{{\rm sn}[2 K(\bar{k})\zeta;\bar{k}]}\right\}\nonumber\\
&=&z^{-1}\sqrt{\frac{2}{u}}\left\{\frac{2 K(\bar{k})\zeta}{{\rm sn}[2 K(\bar{k})\zeta;\bar{k}]}\right\},
\end{eqnarray}
where $\bar{k}^2\ge 0$ is to be determined from
\begin{equation}\label{tkb}
    x_t=-[2K(\bar{k})]^2(\bar{k}^2+1),
\end{equation}
and $\zeta=z/L$, i.e., $0\le\zeta\le 1$.
\end{enumerate}
Here $K(k)$ is the complete elliptic integral of the first kind,
${\rm dn}(\zeta;k)$ and ${\rm sn}(\zeta;k)$ are the Jacobian
delta amplitude and the sine amplitude functions, respectively.
The bulk critical point $T=T_c$ corresponds to $k^2=1/2$.
The above expressions are consistent with the following scaling
form for the order parameter:
\begin{equation}\label{sf}
\phi(z)=L^{-\beta/\nu}X_\phi\left(z/L,tL^{1/\nu}\right),
\end{equation}
with $\beta=\nu=1/2$. Note, however, that within the mean-field theory the magnitude of the scaling function $X_\phi$ is not universal, in that it is multiplied by the nonuniversal factor $\sqrt{2/u}\simeq 1.091$. Finally, we stress that the choice of two parameterizations (see Eqs. (\ref{tk}) and (\ref{tkb})) of the scaling functions in Eqs. (\ref{mz}) and (\ref{mzd}), is just for convenience; it allows one to avoid using imaginary values of $k$ and $\bar{k}$. Indeed, one can transfer any of the set of equations into the other. For example,  defining $\bar{k}$ as
\begin{equation}\label{krel}
\bar{k}=i \frac{k}{k'}, \qquad \text{where} \qquad k'^{\,2}=1-k^2,
\end{equation}
and taking into account the following properties of the elliptic functions \cite{GR73,abramowitz}
\begin{equation}\label{kKrel}
K(\bar{k})=k'\; K(k),
\end{equation}
and
\begin{equation}\label{dnsnrel}
\frac{{\rm dn}(u;ik)}{{\rm sn}(u;ik}=\frac{\sqrt{1+k^2}}{{\rm sn}(u \sqrt{1+k^2};k/\sqrt{1+k^2})},
\end{equation}
one can easily check that the pair of equations (\ref{mz}), (\ref{tk}) is equivalent to the pair of equations (\ref{mzd}), (\ref{tkb}).

\begin{figure}[htbp]
\begin{center}
\includegraphics[width=3in]{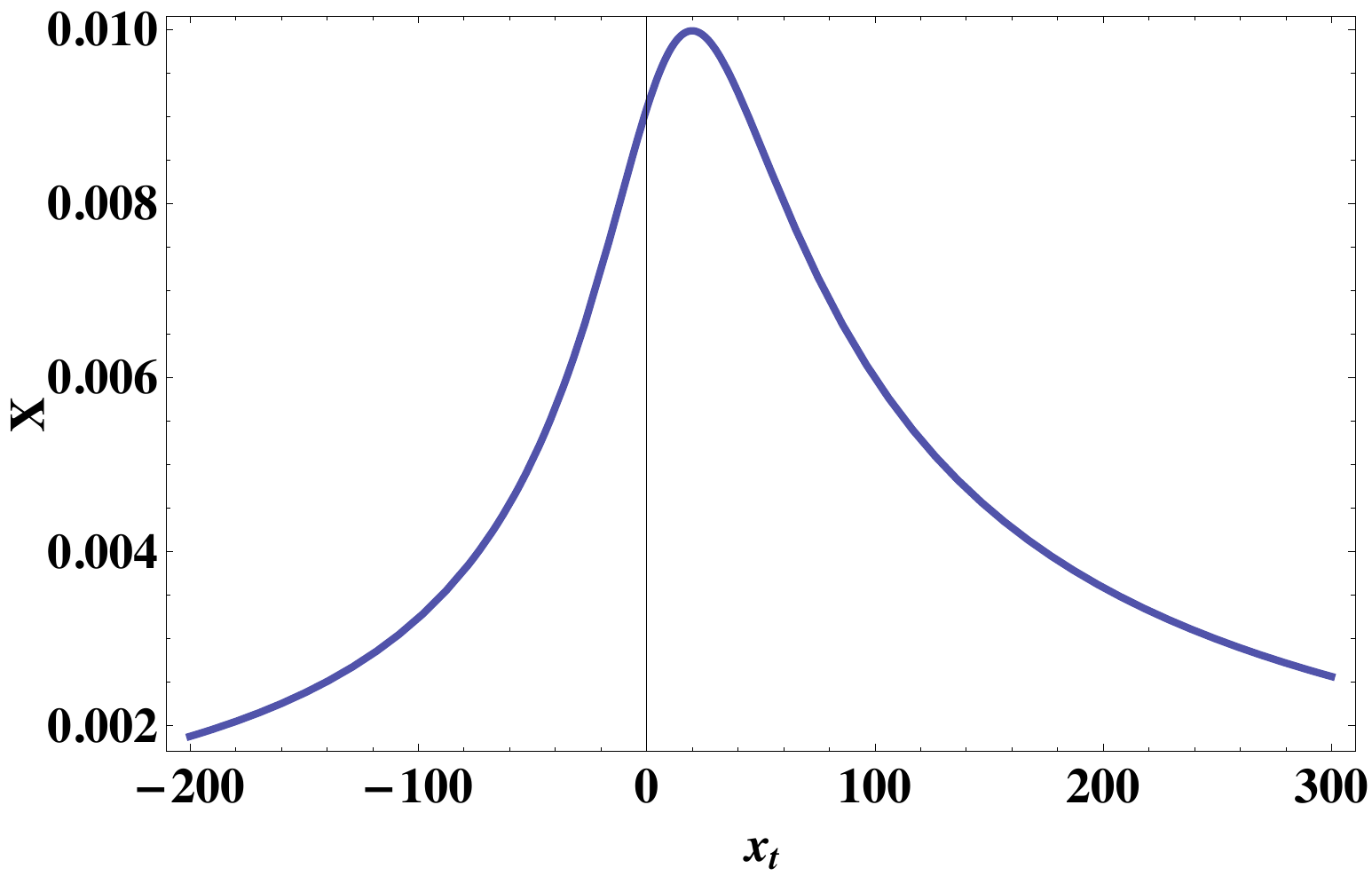}
\caption{Color online. The scaling function of the total susceptibility $X(x_t)$.}
\label{fig:totsuscont}
\end{center}
\end{figure}

In the this article---see Appendix \ref{app:chianalytic}---we report the derivation of exact mean-field expressions for the behavior of the local and of the total susceptibilities. We demonstrate that, when $h=0$, one has
\begin{equation}\label{scafunchilocmt}
\chi_l(z|x_t)=L^{2}X_\chi(z|x_t),
\end{equation}
and
\begin{equation}\label{scafunchimt}
\chi(x_t)=L^{2}X(x_t),
\end{equation}
for the local and the total susceptibilities, respectively. The scaling functions of the total susceptibility $X(x_t)$ is
\begin{equation}\label{Xfinalmt}
X(x_t)=\frac{c_2(x_t)/K\left(k\right)+K\left(k\right)-2 E\left(k\right)}{4
   K^3\left(k\right)},
\end{equation}
where
\begin{equation}\label{c2explmt}
c_2(x_t)=\frac{4 k'^2 k^2 K\left(k\right)}{k'^2
   K\left(k\right)+\left(k^2-k'^2\right) E\left(k\right)}.
\end{equation}
Here $E(k)$ is the complete elliptic integral of the second kind.
The behavior of $X(x_t)$ is illustrated in Fig. \ref{fig:totsuscont}.
For the scaling function $X_\chi(z|x_t)$ of the local susceptibility one has
\begin{equation}\label{Xchigenmt}
X_\chi(\zeta|x_t)=\psi_i (\zeta|x_t)+c_2 \psi_2 (\zeta|x_t),
\end{equation}
where
\begin{equation}\label{partsolexplmt}
\psi_i(\zeta|x_t)=-\frac{k'^{\,2}}{X_{m,0}^2}\left\{1-2\,{\rm dn}\left[i \frac{X_{m,0}}{k'}\,\zeta;k'\right]^2\right\},
\end{equation}
and
\begin{eqnarray}
\lefteqn{\psi_2(\zeta|x_t) = -\frac{k'}{k^2 X_{m,0}^3}\left\{ {\rm dn}\left(i \frac{X_{m,0}}{k'}\,\zeta;k' \right) {\rm sn}\left(i \frac{X_{m,0}}{k'}\,\zeta;k' \right) \right.} \nonumber \\
&& \left[k'(1-2k'^2){\rm E}\left({\rm am}\left(i \frac{X_{m,0}}{k'}\,\zeta;k' \right);k'\right)-ik^2X_{m,0}\zeta\right] \nonumber  \\
&&+k'{\rm cn}\left(i \frac{X_{m,0}}{k'}\,\zeta;k' \right)\\
\times && \left.\left[k'^2+(1-2k'^2)\;{\rm dn}\left(i \frac{X_{m,0}}{k'}\,\zeta;k' \right)2\right]\right\}, \nonumber
\label{secondsolexplmt}
\end{eqnarray}
with
$\zeta\in [-1/2,1/2]$ and
\begin{equation}\label{Xm0defmt}
X_{m,0} \equiv 2 k' K(k)=2 K(\bar{k}).
\end{equation}
\begin{figure}[htbp]
\begin{center}
\includegraphics[width=3in]{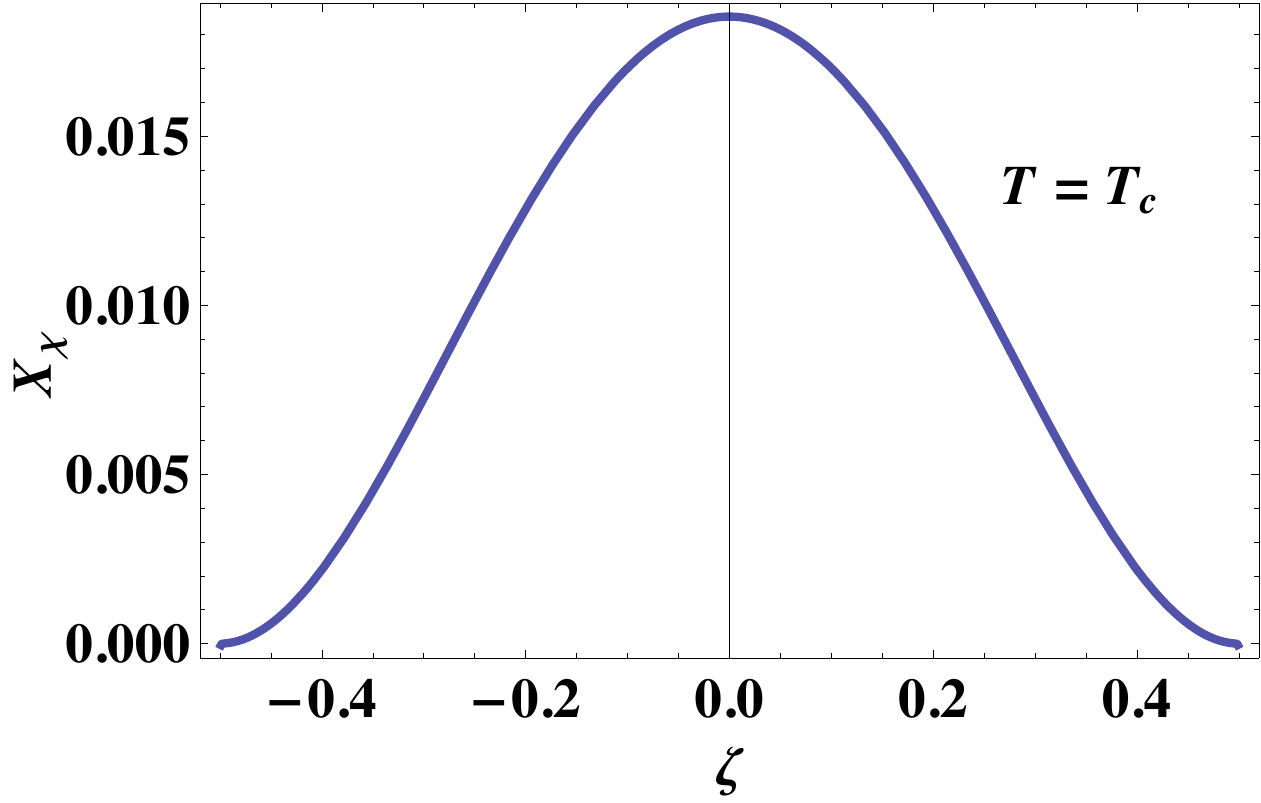}
\caption{Color online. Plot of the function $X_\chi(\zeta|x_t=0)$.}
\label{fig:XchiTc}
\end{center}
\end{figure}
The behavior of the scaling function $X_\chi$ at $T=T_c$ is shown on Fig. \ref{fig:XchiTc}.

\begin{figure}[h!]
\begin{center}
\includegraphics[width=3in]{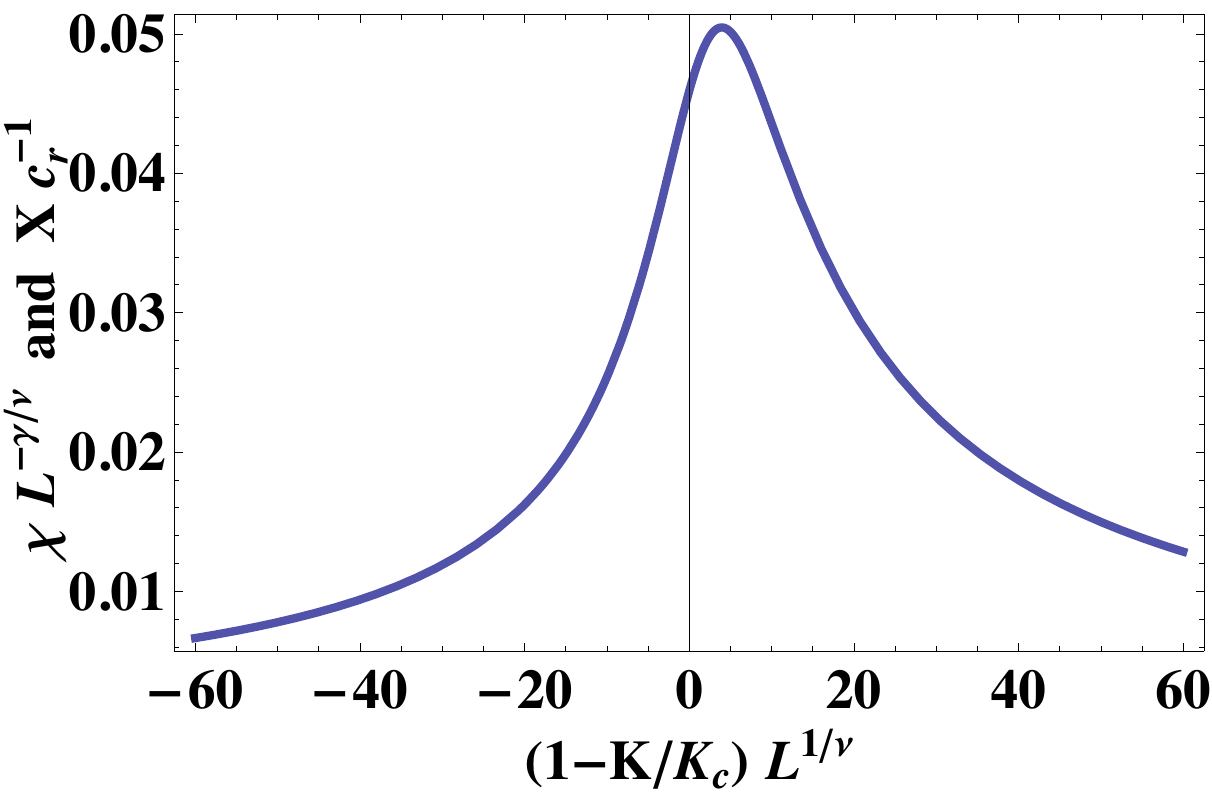}
\caption{Color online. The scaling function of the total susceptibility $\chi L^{-\gamma/\nu}$, calculated via the lattice model, compared versus the scaling function $X(x_t)$ of the same quantity as derived analytically within the continuum approach.}
\label{fig:totsuscontlat}
\end{center}
\end{figure}

As we will see in Appendix \ref{app:chianalytic}, under proper rescaling of the abscissa and the vertical axes that follows from the mapping of the lattice onto the continuum model, there is perfect agreement between our lattice model results for the total susceptibility and the analytically derived ones in the case of a short-ranges system without gravity. The comparison is presented in Fig. \ref{fig:totsuscontlat}, where $c_r=(3c_2^{\rm nn}K_c)^{-1}=0.198$.

\section{The behavior of the susceptibility at $T=T_c$}
\label{at_criticality}

\begin{figure}[htb]
\includegraphics[width=\columnwidth]{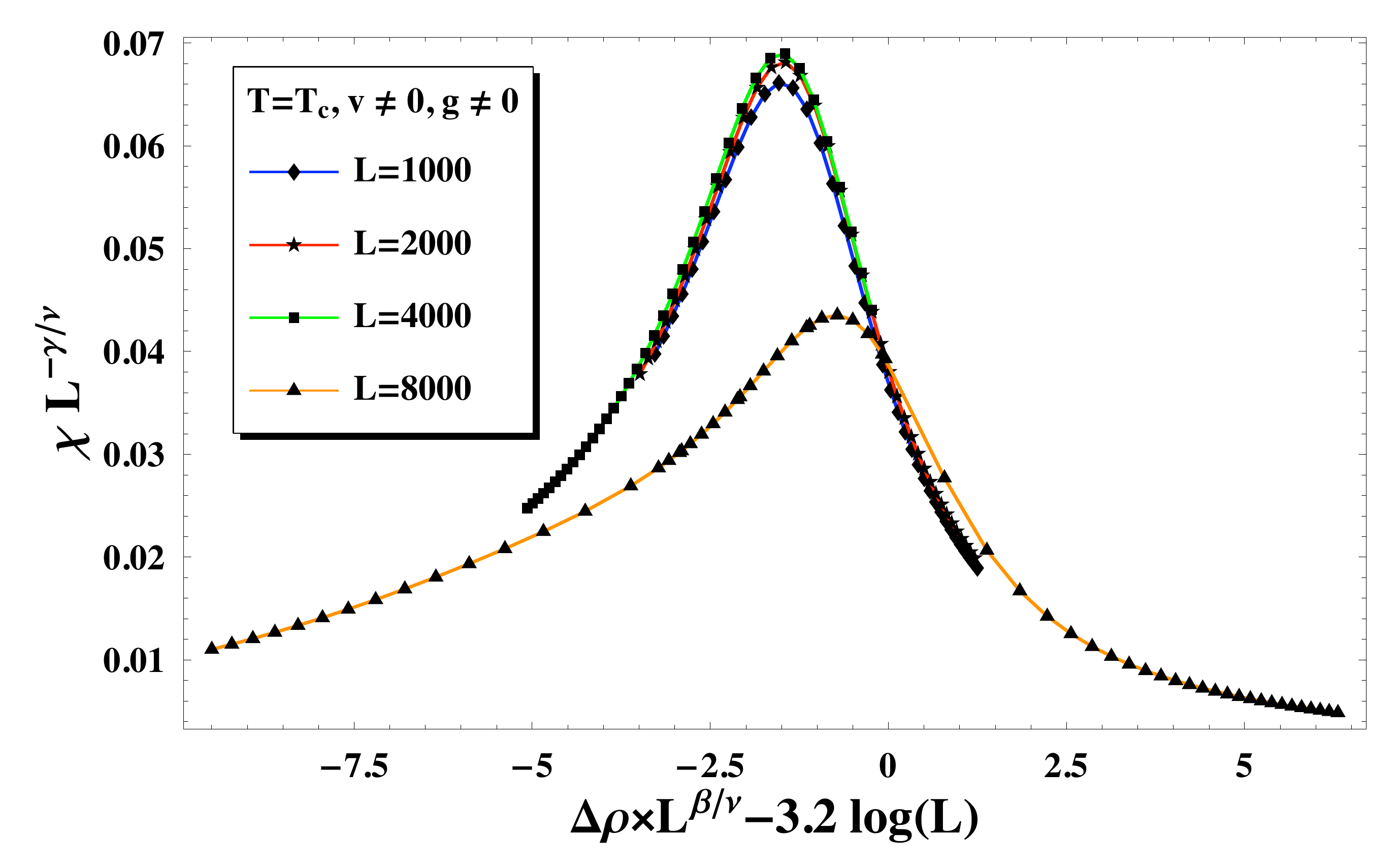}
\caption{Color online. The behavior of the finite-size susceptibility $\chi$ at $T=T_c$ as a function of
$x_\rho=\Delta\rho L^{\beta/\nu}-3.2 \ln L$ for films with thickness $L=1000, 2000, 4000$ and $8000$
layers. Both van der Waals forces and gravity effects are taken into account.
  \label{chi_verus_rho_all}}
\end{figure}

\begin{figure}[htb]
\includegraphics[width=\columnwidth]{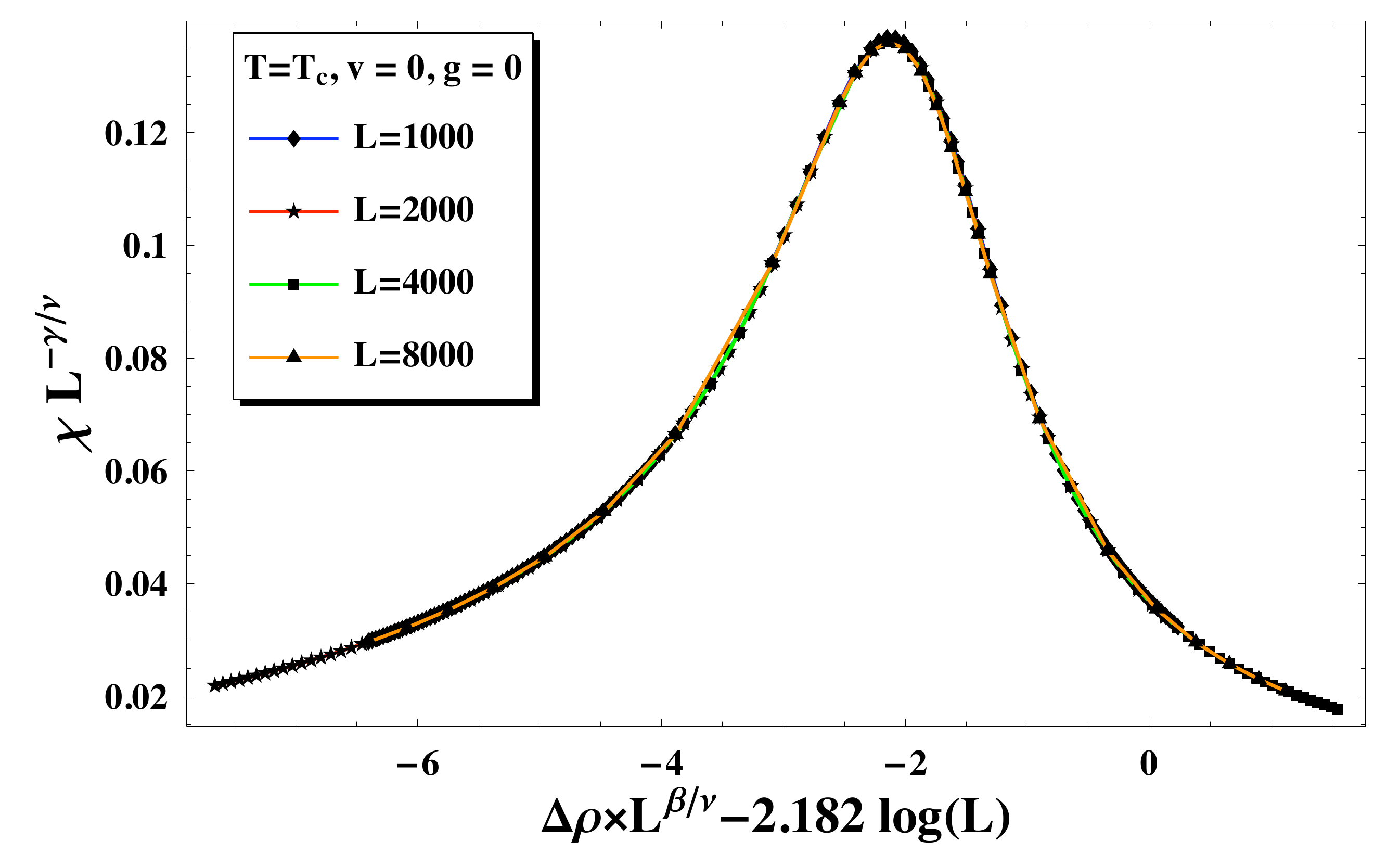}
\caption{Color online. The behavior of the finite-size susceptibility $\chi$ at $T=T_c$ as a function of
$x_\rho=\Delta\rho L^{\beta/\nu}-2.182 \ln L$ for films with thickness $L=1000, 2000, 4000$ and $8000$
layers for systems with
only short-ranged interaction present.
  \label{chi_verus_rho_sr_all}}
\end{figure}

\begin{figure*}[htb]
\begin{center}
\begin{tabular}{cc}
\resizebox{\columnwidth}{!}{\includegraphics{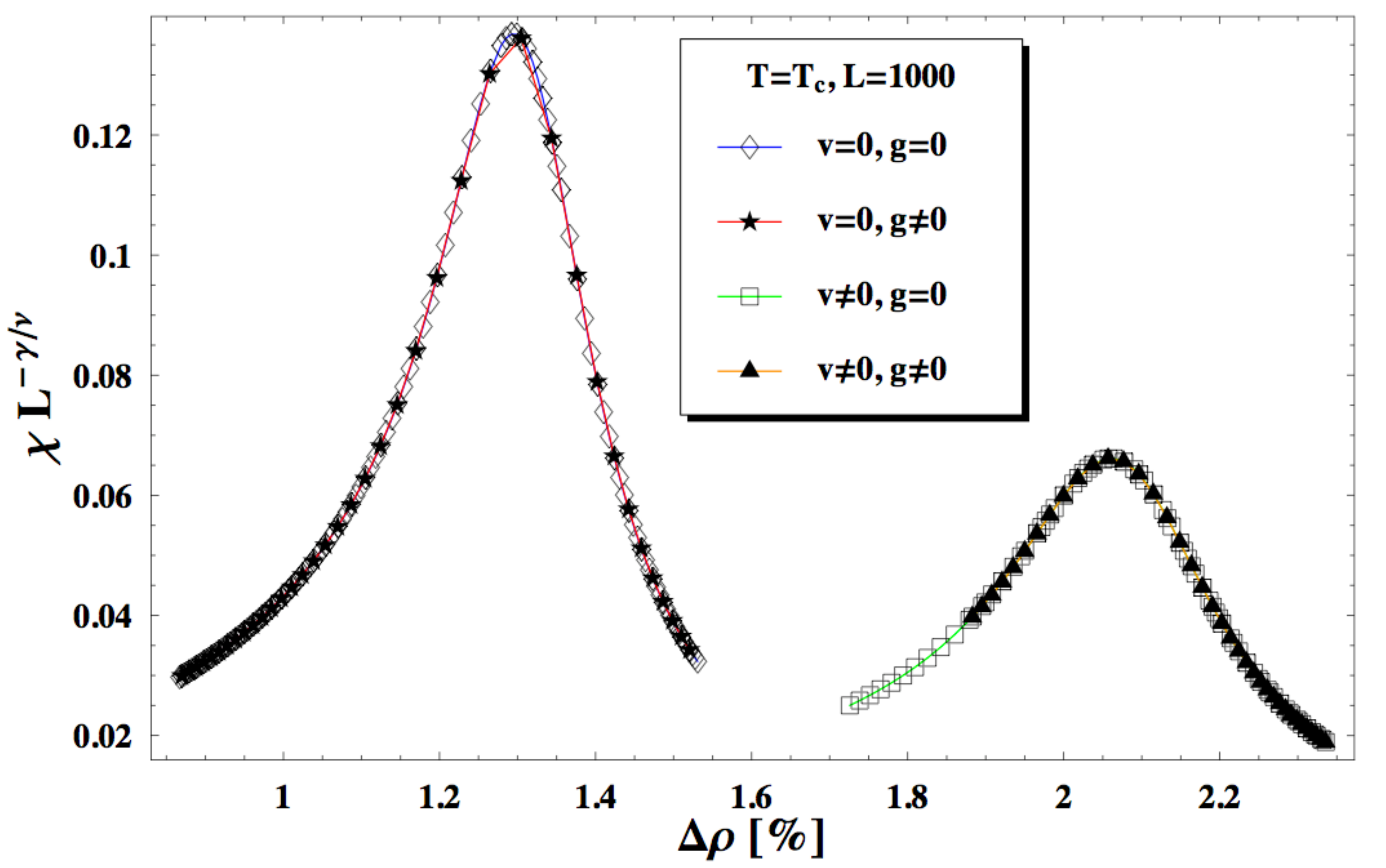}}
& \resizebox{\columnwidth}{!}{\includegraphics{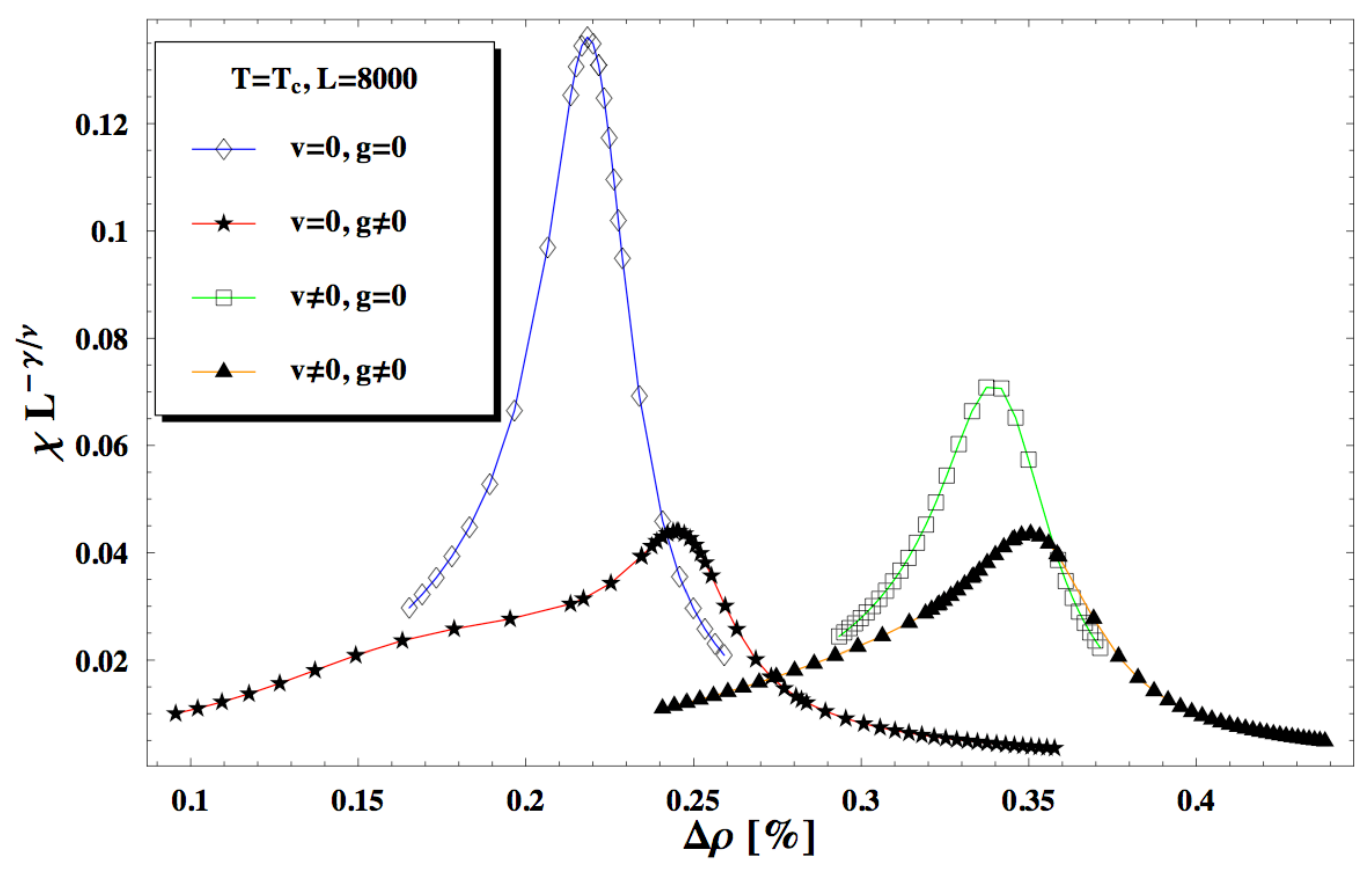}}
\end{tabular}
\end{center}
\caption{Color online. The behavior of the finite-size susceptibility $\chi$ at $T=T_c$ as a function
of $\Delta\rho$ as obtained within the mean-field like treatment of the model.
All the curves are for $L=1000$ or $L=8000$  layers thick film but with  gravity
and/or van der Waals interaction neglected or taken into account.
One observes that both van der Waals interactions, as well as gravity,
suppress the divergence of the susceptibility.
  \label{chi_verus_rho_8000}}
\end{figure*}

\begin{figure*}[htb]
\begin{center}
\begin{tabular}{cc}
\resizebox{\columnwidth}{!}{\includegraphics{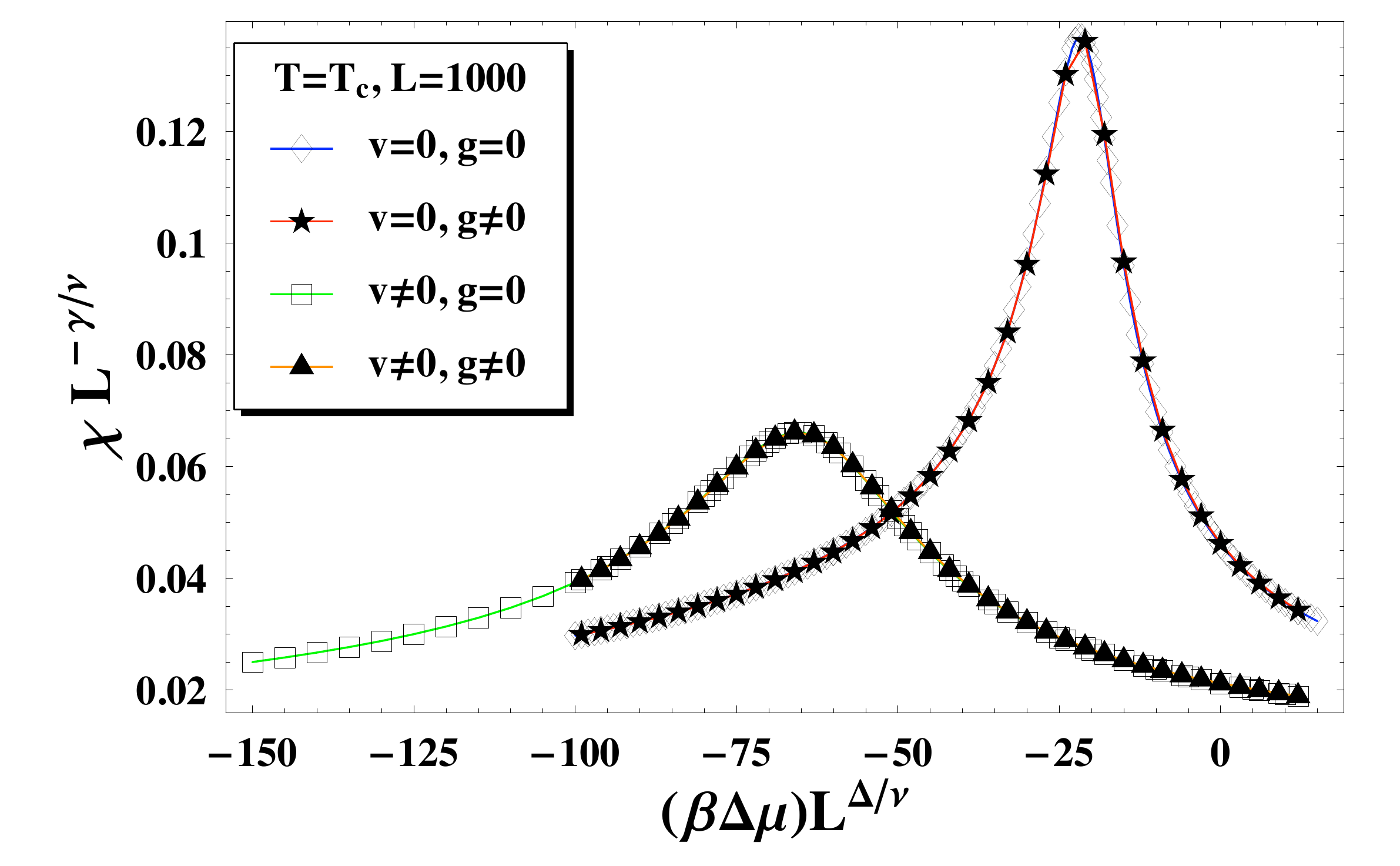}}
& \resizebox{\columnwidth}{!}{\includegraphics{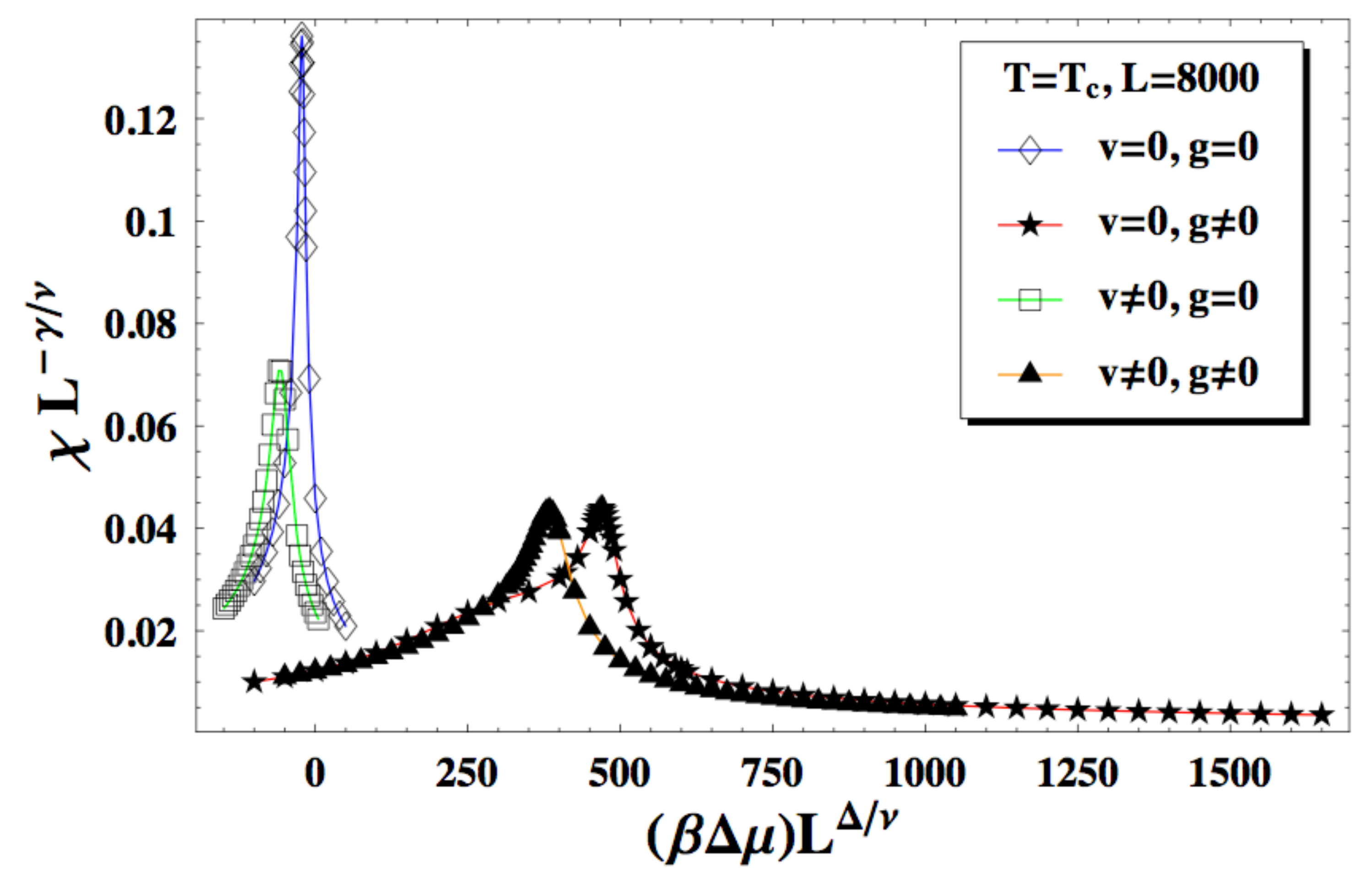}}
\end{tabular}
\end{center}
\caption{Color online. The same as in Fig. \ref{chi_verus_rho_8000} but now the behavior of
the finite-size susceptibility $\chi$ is considered as a function of the
scaling variable $(\beta\Delta\mu) L^{\Delta/\nu}$.
  \label{chi_verus_mu_8000}}
\end{figure*}

\begin{figure*}[htb]
\begin{center}
\begin{tabular}{cc}
\resizebox{\columnwidth}{!}{\includegraphics{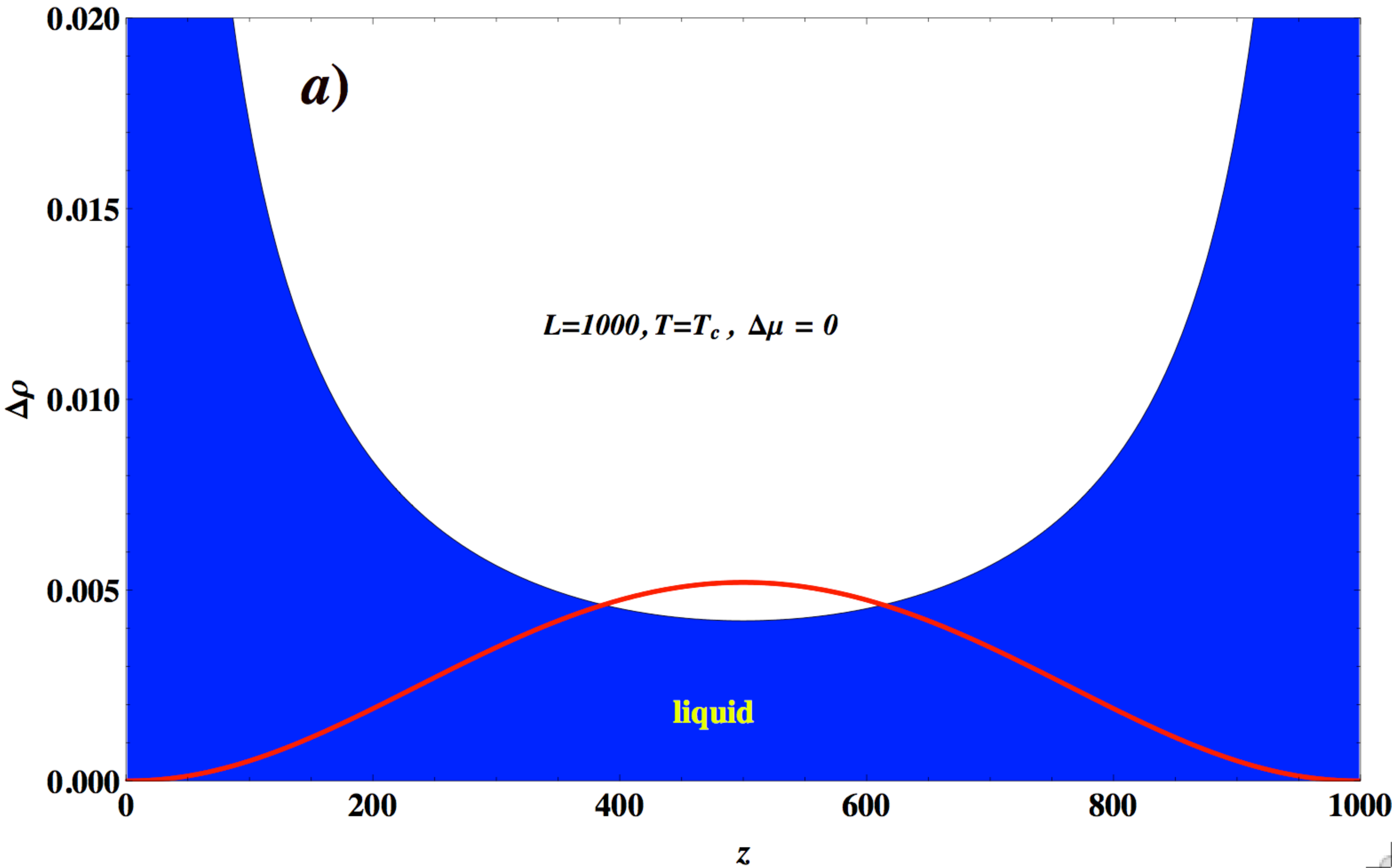}} &
\resizebox{\columnwidth}{!}{\includegraphics{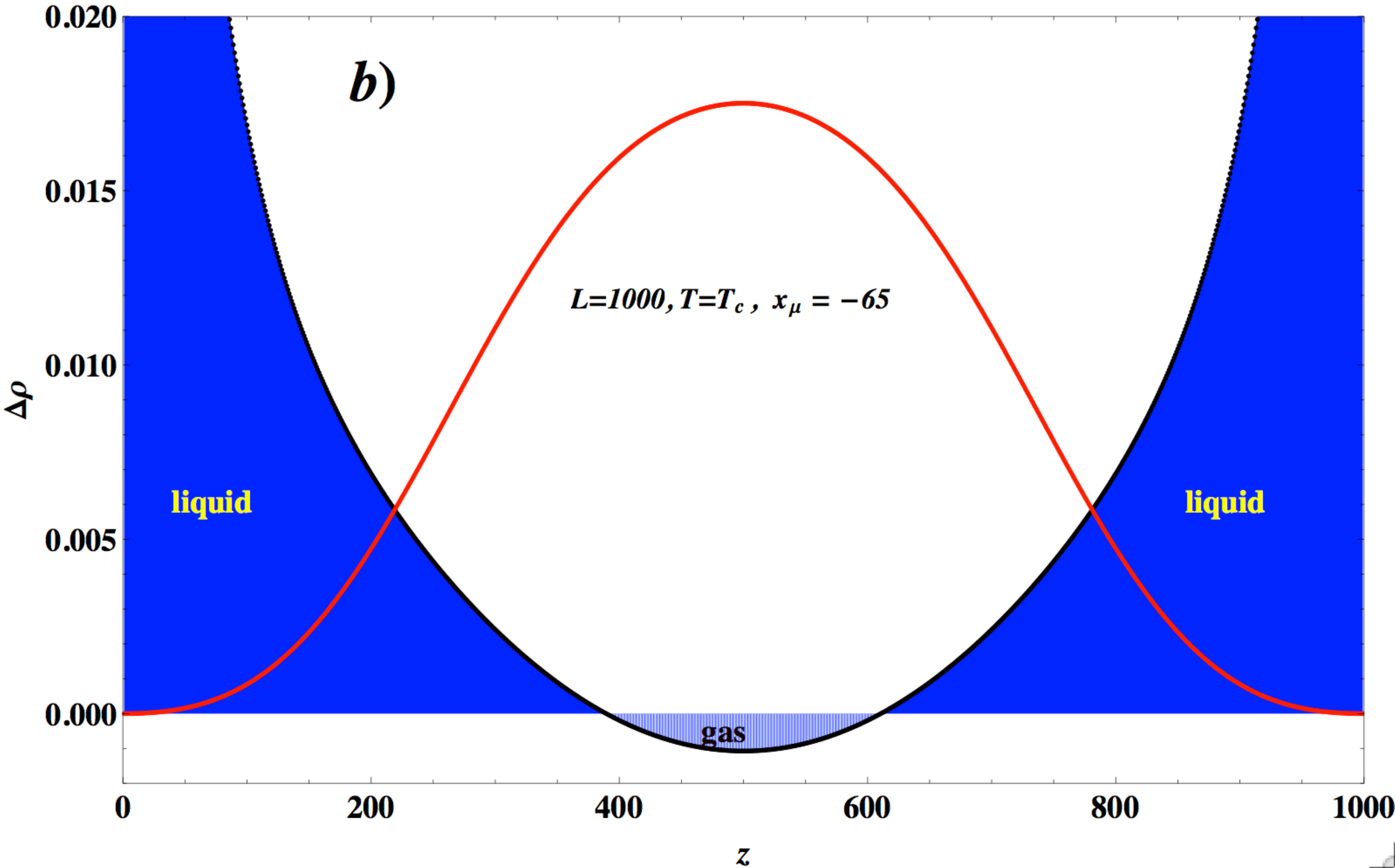}} \\[0.5cm]
\resizebox{\columnwidth}{!}{\includegraphics{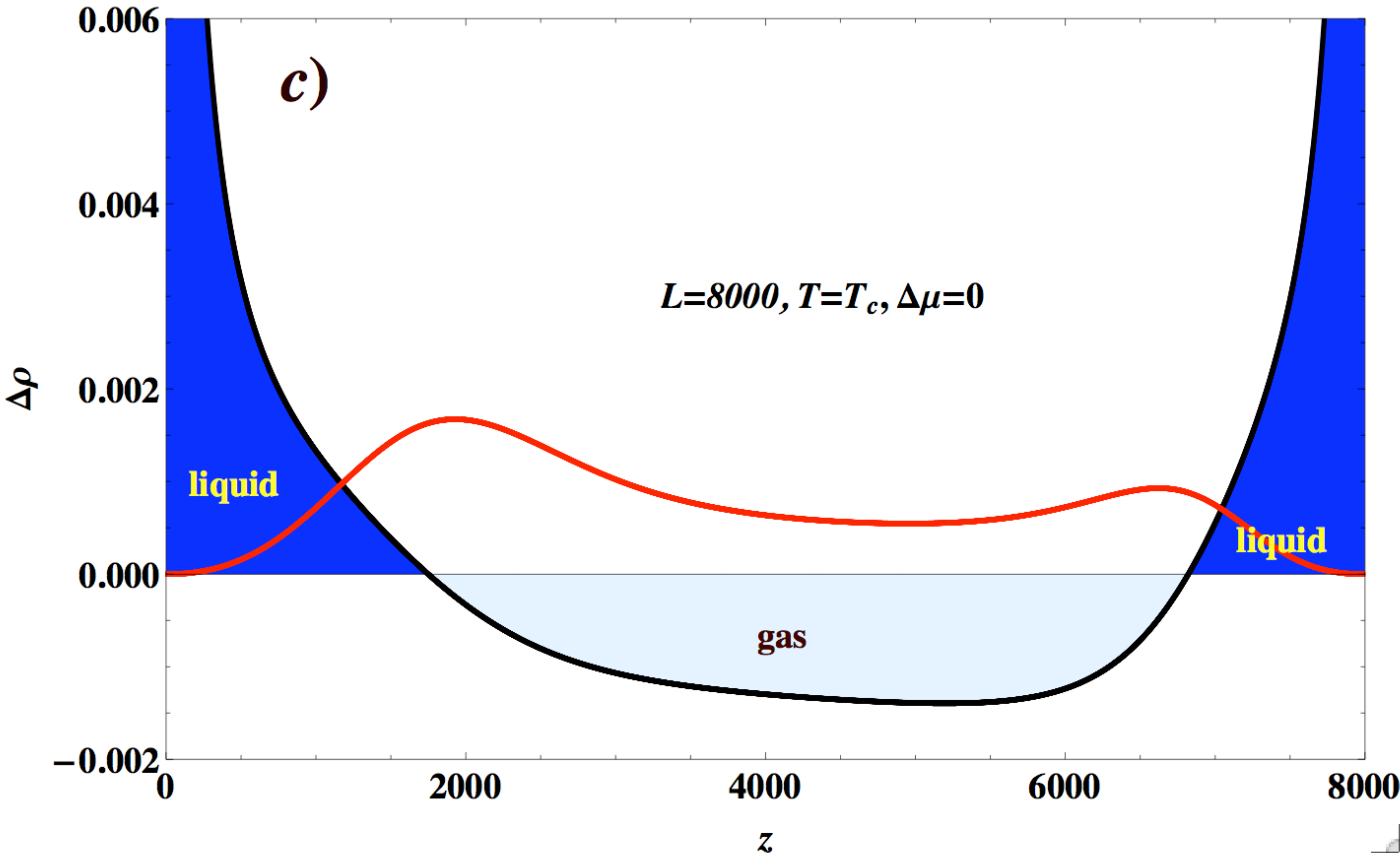}} &
\resizebox{\columnwidth}{!}{\includegraphics{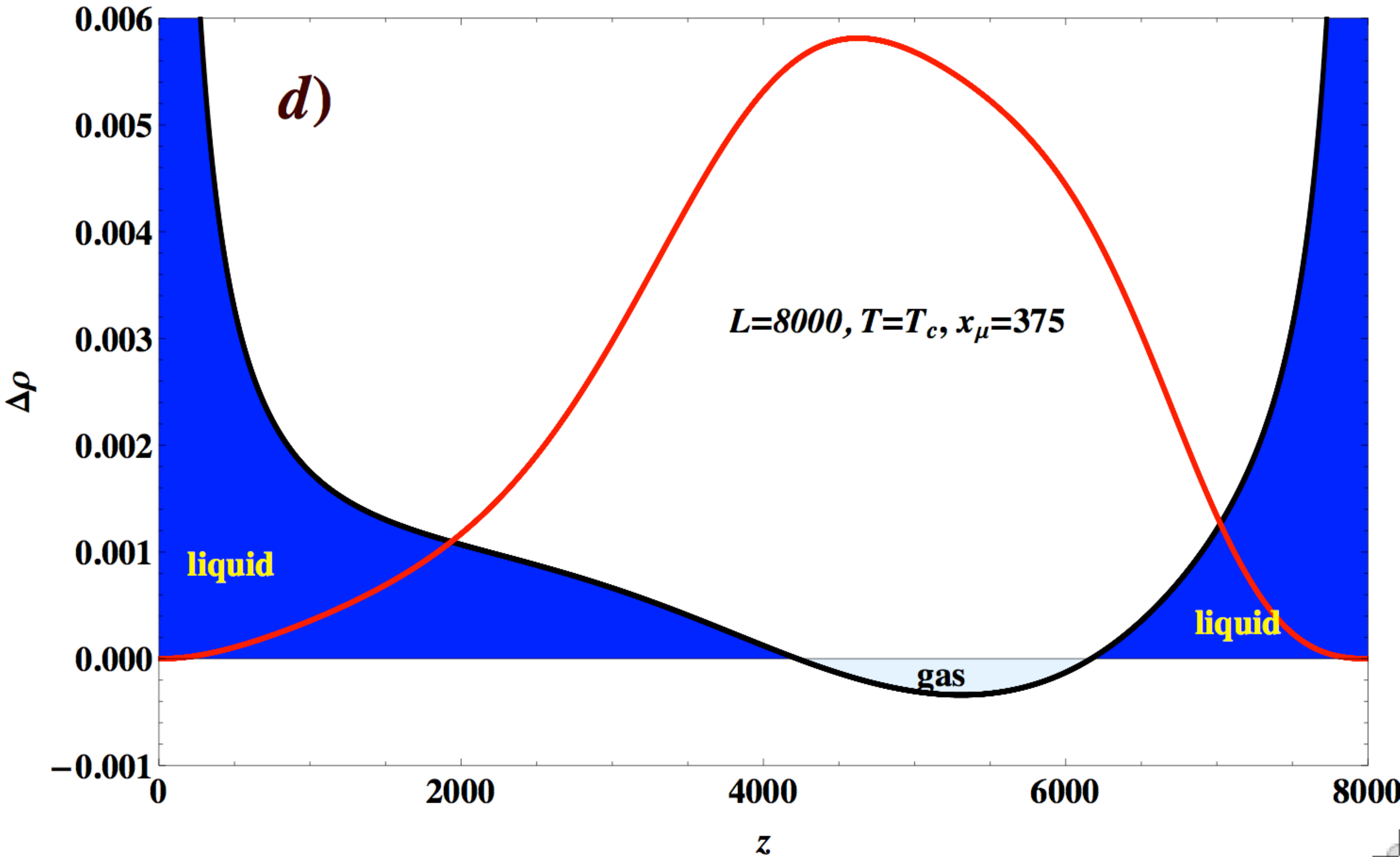}}
\end{tabular}
\end{center}
\caption{Color online. a) The density profile for a system with $L=1000$ layers at $T=T_c$ and $\mu=\mu_c$.
b) The density profile for a system with $L=1000$ layers at $T=T_c$
and $\Delta\mu=\Delta\mu_{\rm max}$ for which the susceptibility
reaches its maximum.
c) The density profile for a system with $L=8000$
layers at $T=T_c$ and $\mu=\mu_c$.
d) The density profile for a system with $L=8000$ layers
at $T=T_c$ and $\Delta\mu=\Delta\mu_{\rm max}$  for which the susceptibility
reaches its maximum. In all the figures the bold (red) curve shows the profile of the local susceptibility normalized by $L^2$ for the system with size $L$. In order to have a better visibility of the order parameter profile simultaneously with the profile of the local susceptibility, the magnitude of the local susceptibility has been further reduced $8$ times for $L=1000$ and $16$ times for $L=8000$.
} \label{density}
\end{figure*}

Our analysis of the finite-size behavior of the system   at $T=T_c$ is summarized  in Figs.
\ref{chi_verus_rho_all}, \ref{chi_verus_rho_sr_all},
 \ref{chi_verus_rho_8000},
\ref{chi_verus_mu_8000}, and \ref{density}. Specifically, Fig. \ref{chi_verus_rho_all} presents
the behavior of the normalized susceptibility $\chi L^{-\gamma/\nu}$ in an experimentally
realistic system in which both gravity and van der Waals interactions are taken into account,
while Fig. \ref{chi_verus_rho_sr_all} shows the behavior of the susceptibility in one idealized
system in which only short-ranged interactions are taken into account.
In both figures the behavior of the finite-size susceptibility
is shown as a function of
$x_\rho=\Delta\rho L^{\beta/\nu}- c \ln L$, where $c=2.182$ for the short-ranged systems and $c=3.2$ for systems with van der Waals interaction present. Films with thickness $L=1000, 2000, 4000$ and $L=8000$
layers are considered,    and
 \begin{equation}\label{deltarho}
 \Delta \rho =\frac{1}{L}\sum_{z=0}^L \phi(z).
 \end{equation}

The appearance of the $\ln L$ corrections in the scaling variable $x_\rho$ is a specific feature of the mean-field systems due to the degeneracy of the  critical exponents $\beta$ and $\nu$ which become equal in mean-field approximation. The numerical value of the constant $c$ can be analytically predicted: at the bulk critical point ${K=K_c=(c_2+2c_2^{\rm nn})^{-1}\simeq 0.168,\hat{h}=0}$ one has $a=0$, $u=1.681$  and Eq. (\ref{eqcontsrmt}) possesses a solution, see Eq. (\ref{mz}), where the leading behavior near the boundary, when $\zeta\to 0$, is
\begin{equation}\label{simsol}
\phi(z)\simeq z^{-1}\sqrt{2/u}\;\simeq 1.091/z.
\end{equation}
Integrating over $z$ and having in mind that the system is bounded by two substrate planes, one immediately obtains a $\ln L$ contribution in $\Delta \rho$ which is proportional to $c=2.182$, i.e., exactly the same constant as given in Fig.~\ref{chi_verus_rho_sr_all}.  The presence of a van der Waals type interaction changes the constants of the model, e.g. the critical coupling $K_c$, as well as the effective constant in front of the second derivative of the order parameter profile. The last does not change the leading $z$-dependence of the order parameter profile but leads to a different constant $c$, which turns out to be $c=3.2$ in our model. One observes that when both van der Waals forces and gravity effects
are taken into account, see Fig.\ref{chi_verus_rho_all},  but $L$ is not very large,
the gravity effects are negligible - as for $L=1000, 2000$ and $4000$; the corresponding
curves are close to each other and one can speak
about (some) data collapse for them. However, the curve for a system with $L=8000$, for which
the gravity effects are essential, differs essentially from the others  with the maximum of the
susceptibility strongly suppressed. Furthermore, note that in the presence of van der Waals interactions
and gravity the maximum of the curve for $L=8000$ shifts to higher
values of $x_\rho$ in comparison with systems with short-ranged interactions
only and no gravity; see Fig. \ref{chi_verus_rho_sr_all}.  In the last case  one observes a
reasonable data collapse for all values of $L$ considered. We stress that $\Delta\rho$
contains contributions due to the role of the boundaries which are not,
in fact, critical. These  are, e.g., the contributions which are due to the
layers very near the boundaries,  which layers are, independently on the value of $x_\rho$,
always liquid like and almost fully occupied (i.e. densely packed) by the molecules of the fluid.
Thus, in terms of $\Delta\rho$ one expects larger corrections to scaling
than when $\chi$ is considered as a function of $\Delta\mu$ (see Fig. \ref{chi_verus_mu_sr_all}).
The behavior of the finite-size susceptibility $\chi$ at $T=T_c$ as a function
of $\Delta\rho$, as obtained within the mean-field like treatment of the model, is shown in Fig.
\ref{chi_verus_rho_8000}.
All the curves are for $L=1000$ or $L=8000$  layers thick film but with  gravity
and/or van der Waals interaction neglected or taken into account.
One observes that both van der Waals interactions, as well as gravity,
suppress the divergence of the susceptibility. Fig. \ref{chi_verus_mu_8000} shows the same as
Fig. \ref{chi_verus_rho_8000} but now the behavior of
the finite-size susceptibility $\chi$ is considered as a function of the
scaling variable $(\beta\Delta\mu) L^{\Delta/\nu}$.
One observes the different importance of the van der Waals
substrate-fluid interactions and of the gravity within ``thin" films,
exemplified by $L=1000$, and ``thick" films, represented by the case $L=8000$.
While for $L=1000$ the curves group together depending on the presence or absence
of van der Waals substrate-fluid interactions, for $L=8000$ they do this,
at least with respect to the position of their maximum, depending on the presence of gravity
in the system. Furthermore, one observes that in ``thick" films when $g=0$
the maximum of the susceptibility is at $\Delta\mu <0$, {\it contrary}
to the case with $g\ne 0$, when the maximum is at $\Delta\mu >0$.
In ``thin" films the position of the maximum does not depend on $g$
and always is at $\Delta\mu <0$. Finally, Fig. \ref{density} shows the density
profiles of systems with $L=1000$ and $L=8000$ layers both for $x_\mu=0$, as
well as for the corresponding $x_\mu$ for which the susceptibility has a maximum.
More precisely, the part a) of this figure shows the density
profile for a system with $L=1000$ layers at $T=T_c$ and $\mu=\mu_c$.
Note that $\Delta\rho>0$ everywhere, i.e., the equilibrium state of the film
is {\it liquid}-like. For $L=1000$ the effect of gravity is negligible, i.e.,
this profile is similar to the one with for $L=8000$ when $g=0$.
Part b) of Fig. \ref{density} shows the density profile for a system with $L=1000$
layers at $T=T_c$
and $\Delta\mu=\Delta\mu_{\rm max}$ for which the susceptibility
reaches its maximum. Note that $\Delta\mu_{\rm max}<0$. Next, part
c) of Fig. \ref{density} presents the density profile for a system with $L=8000$
layers at $T=T_c$ and $\mu=\mu_c$. Note that, due to the gravity,
the profile is with $\Delta\rho>0$ near the walls, but $\Delta\rho<0$
everywhere in the middle of the system, i.e. in the middle of the system
the equilibrium profile for the finite film is {\it gas}-like.
This has to be compared with case a) when $\Delta\rho>0$ for all $0\leq z\leq L$.
Finally, part d) of Fig. \ref{density} presents the density profile for a system with $L=8000$ layers
at $T=T_c$ and $\Delta\mu=\Delta\mu_{\rm max}$  for which the susceptibility
reaches its maximum. Note that, contrary to the case $L=1000$, when the gravity is
not important,  $\Delta\mu_{\rm max} > 0$.

\section{The behavior of the susceptibility for $T>T_c$ and  $T<T_c$}
\label{off_criticality}

\begin{figure*}[htb]
\begin{center}
\begin{tabular}{cc}
\resizebox{\columnwidth}{!}{\includegraphics{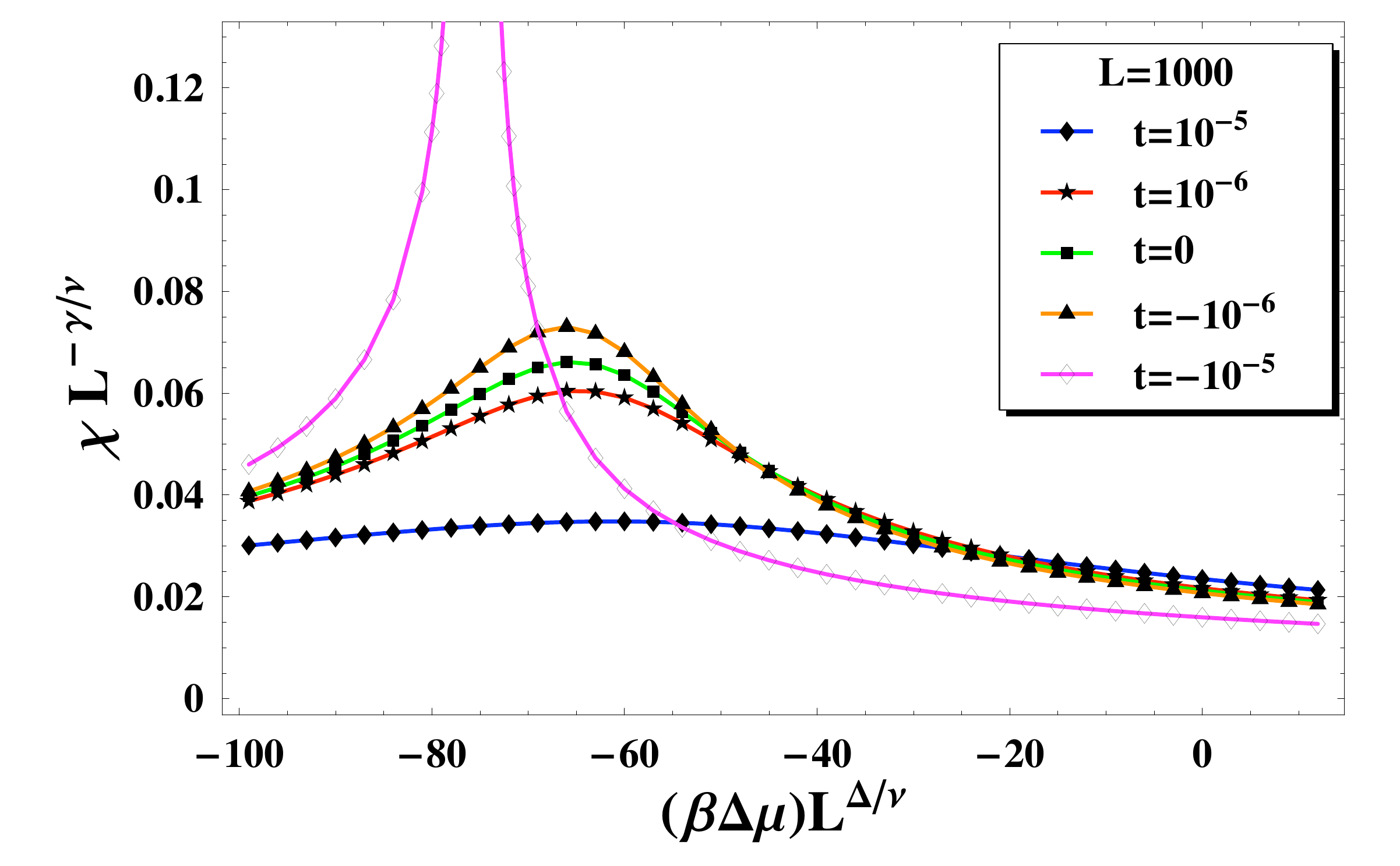}}
& \resizebox{\columnwidth}{!}{\includegraphics{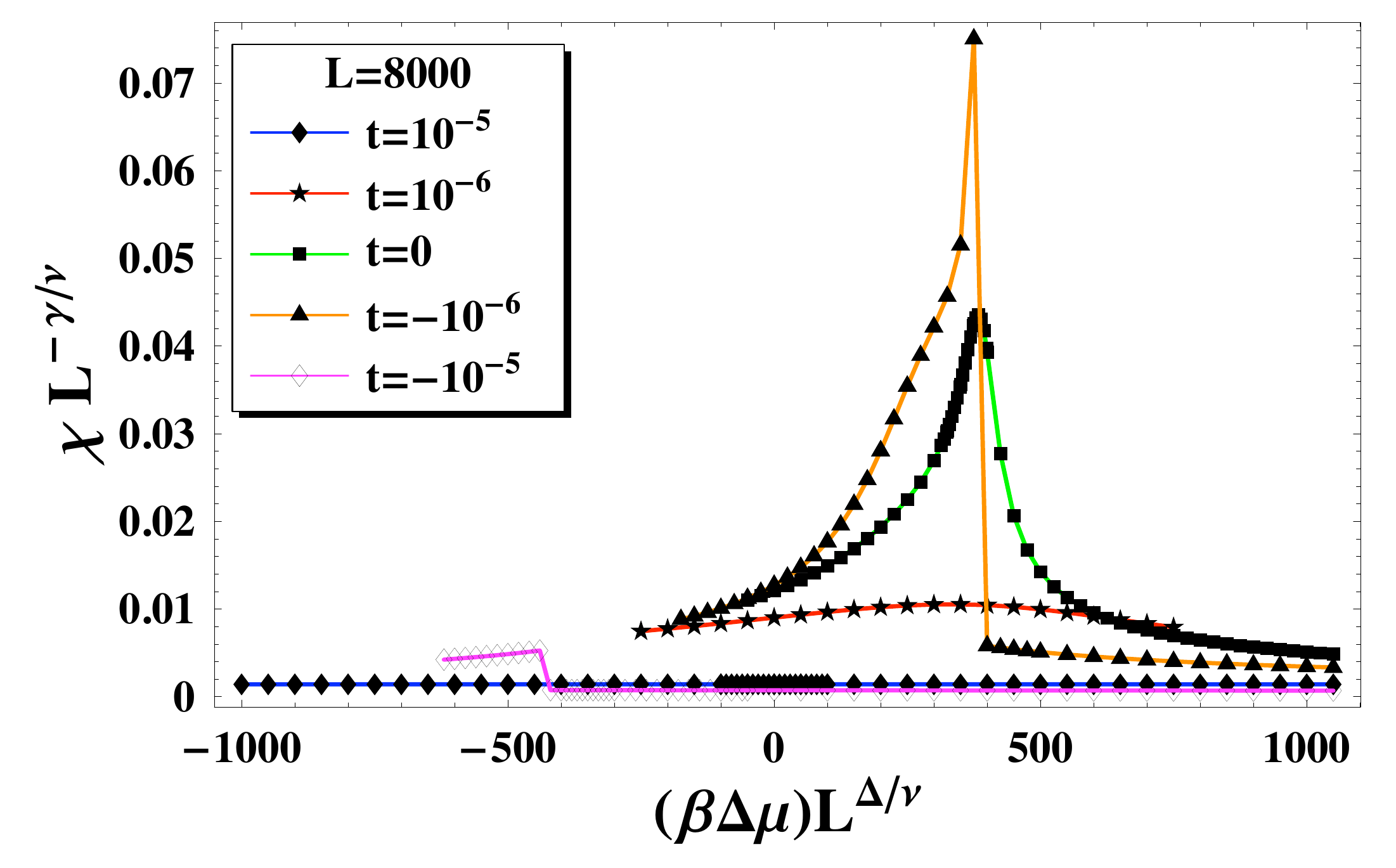}}
\end{tabular}
\end{center}
\caption{Color online. The behavior of the finite-size susceptibility as a function of the scaling
variable $x_\mu=\Delta\mu L^{\Delta/\nu}$ for a film with thicknesses $L=1000$ and
$L=8000$ for $t=0, t=\pm 10^{-6}$ and $t=\pm 10^{-5}$.
  \label{chi_verus_mu_1000}}
\end{figure*}

\begin{figure}[htb]
\includegraphics[width=\columnwidth]{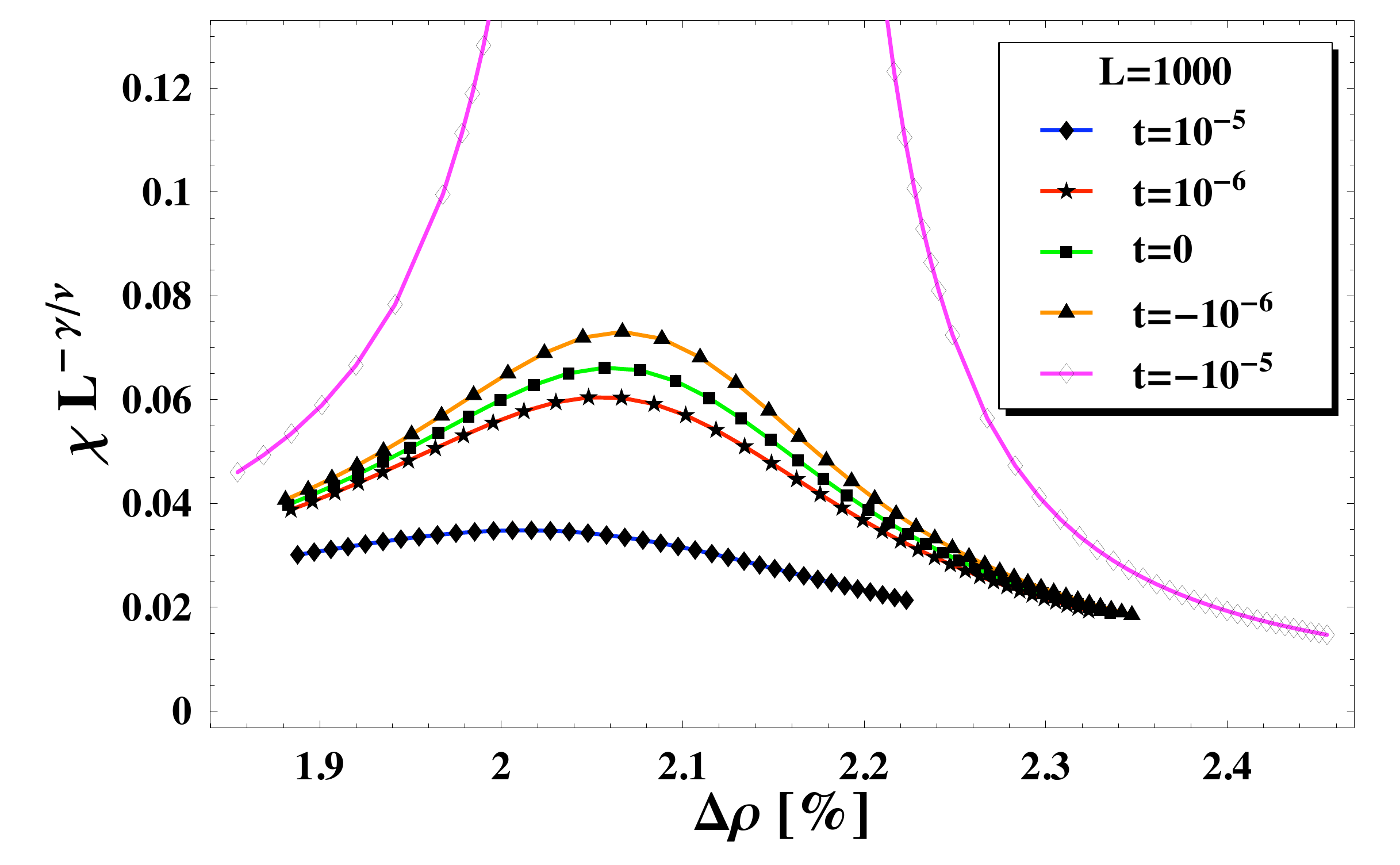}
\caption{Color online. The behavior of the finite-size susceptibility as a function of
$\Delta\rho$ for a film with thickness $L=1000$ and for
$t=0, t=\pm 10^{-6}$ and $t=\pm 10^{-5}$.
  \label{chi_verus_rho_1000}}
\end{figure}

\begin{figure*}[htb]
\begin{center}
\begin{tabular}{cc}
\resizebox{\columnwidth}{!}{\includegraphics{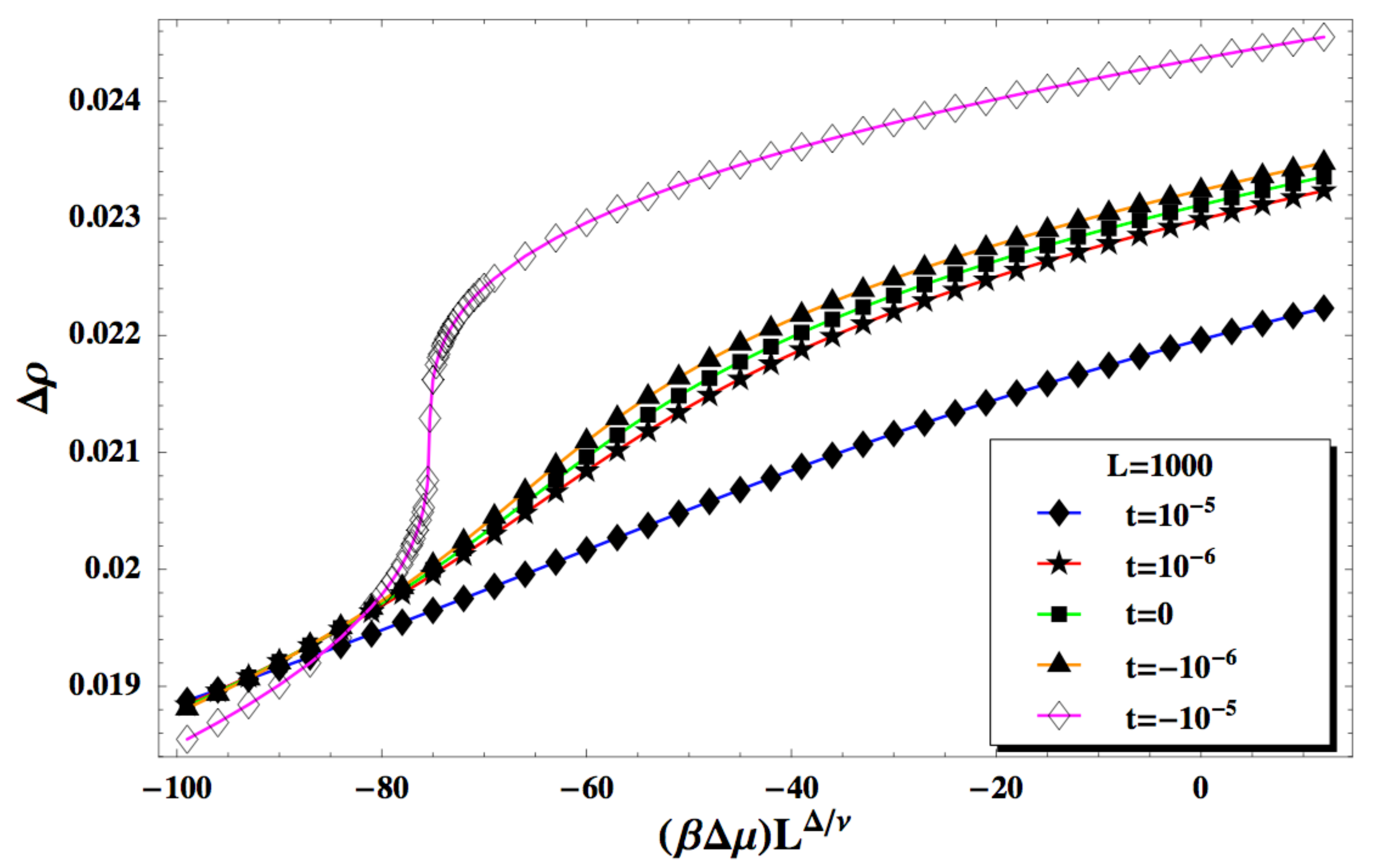}}
& \resizebox{\columnwidth}{!}{\includegraphics{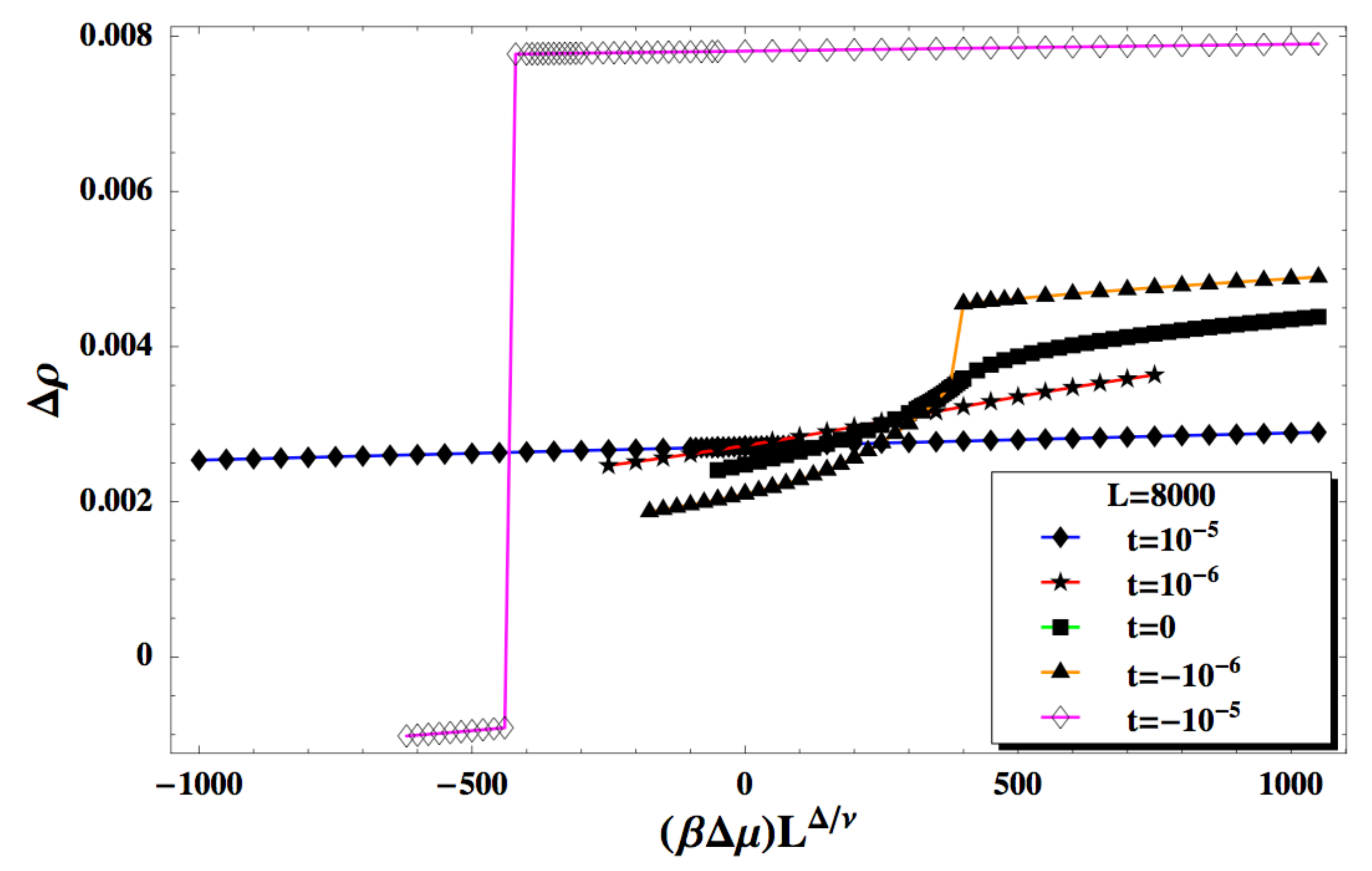}}
\end{tabular}
\end{center}
\caption{Color online. The behavior of $\Delta\rho$ as a function of the scaling variable
$x_\mu=\Delta\mu L^{\Delta/\nu}$ for a film with thicknesses
$L=1000$ and $L=8000$ for $t=0, t=\pm 10^{-6}$ and $t=\pm 10^{-5}$.
  \label{rho_verus_mu_1000}}
\end{figure*}

The behavior of the susceptibility away from the critical point is summarized in Figs.
\ref{chi_verus_mu_1000}, \ref{chi_verus_rho_1000} and \ref{rho_verus_mu_1000}. More precisely,
Fig. \ref{chi_verus_mu_1000} presents the behavior of the finite-size susceptibility as a
function of the scaling
variable $x_\mu=\Delta\mu L^{\Delta/\nu}$ for a film with thicknesses $L=1000$ and
$L=8000$ for $t=0, t=\pm 10^{-6}$ and $t=\pm 10^{-5}$. When
$t=-10^{-5}$ and at a negative value of $x_\mu$ the system undergoes a
second order phase transition for $L=1000$, while for $L=8000$
it undergoes first order phase transitions for both $t=-10^{-6}$ and $t=-10^{-5}$. We further note that when $L=1000$, the gravity effects being negligible, the critical point $T_c(L)$ of the finite system is at  $\Delta \mu_{c,L}<0$, see Fig. \ref{phasediagram}. In general it is quite difficult to determine  the location of this point with good precision \cite{BL91}. Inspecting Fig. \ref{chi_verus_mu_1000} one discovers, however, that for $L=1000$ the curve with $t=-10^{-5}$  is very close to to the corresponding one that characterizes the behavior of the system at the true critical temperature of the finite system.
In the context of the behavior displayed in this figure, it is useful to refer to Fig. \ref{rho_verus_mu_1000}, which illustrates the behavior of the excess normalized density
$\Delta \rho$ as a function of the scaling variable $x_\mu=\Delta\mu L^{\Delta/\nu}$ for the
same values of $L$ and for the same fixed values of $t$ as chosen in Fig. \ref{chi_verus_mu_1000}.

The behavior of the finite-size susceptibility as a function of
$\Delta\rho$ for a film with thicknesses $L=1000$ and for
$t=0, t=\pm 10^{-6}$ and $t=\pm 10^{-5}$ is shown in Fig. \ref{chi_verus_rho_1000}. Finally, the
behavior of $\Delta\rho$ as a function of the scaling variable
$x_\mu=\Delta\mu L^{\Delta/\nu}$ for a film with thicknesses
$L=1000$ and $L=8000$ for $t=0, t=\pm 10^{-6}$ and $t=\pm 10^{-5}$ is shown in Fig.
\ref{rho_verus_mu_1000}.
Again, one observes that when $L=1000$ and $t=-10^{-5}$  the system undergoes a second order phase transition at a negative value
of $x_\mu$ while when $L=8000$ it undergoes a first order transition for both $t=-10^{-5}$
and $t=-10^{-6}$ as $x_\mu$ is varied. We note that near $T_c$ the position of the vapor-liquid coexistence line shifts  (see Fig. \ref{phasediagram}) when $L$ increases,  from a position characterized by $\Delta \mu<0$ to a position with $\Delta \mu>0$. The same is also true for the position of the true critical point of the finite system. For $L=8000$ the critical point $T_c(L=8000)$ is at $\Delta \mu_{c,L}>0$. The larger $L$ the narrower the region in $t$ and $\Delta \mu$ in which one has rounding of the second-order phase transition around the bulk critical point. Furthermore, we note that the larger $L$ is stronger the gravity effects will be in the system; this in turn stimulates the phase separation within the system.

These general remarks are intended to provide the reader with guidance regarding general trends of behavior as the system size $L$ varies between 1000 and 8000. We also stress that the effects of  van der Waals interactions, as well as those of gravity are not universal, in that they depend on the strength of two parameters: the value of the effective surface potential $h_{w,s}$ and the value of the effective gravity constant $g$; see Appendix \ref{gravity}. Thus,  detailed predictions for the finite-size behavior of the susceptibility in a particular fluid bounded by specific substrates require that one perform numerical calculations following the general prescriptions presented in this article.

\section{Discussion and concluding remarks}
We have studied the behavior of the finite-size susceptibility in fluid nonpolar films governed by van der Waals interactions and subjected to the influence of the gravitational field of the Earth.  We focused on the situation in which the film is bounded by solid substrate plane boundaries which both strongly prefer the liquid phase of the fluid (i.e. we considered the so-called ``plus-plus'' boundary condition).

\begin{figure}[htb]
\includegraphics[width=\columnwidth]{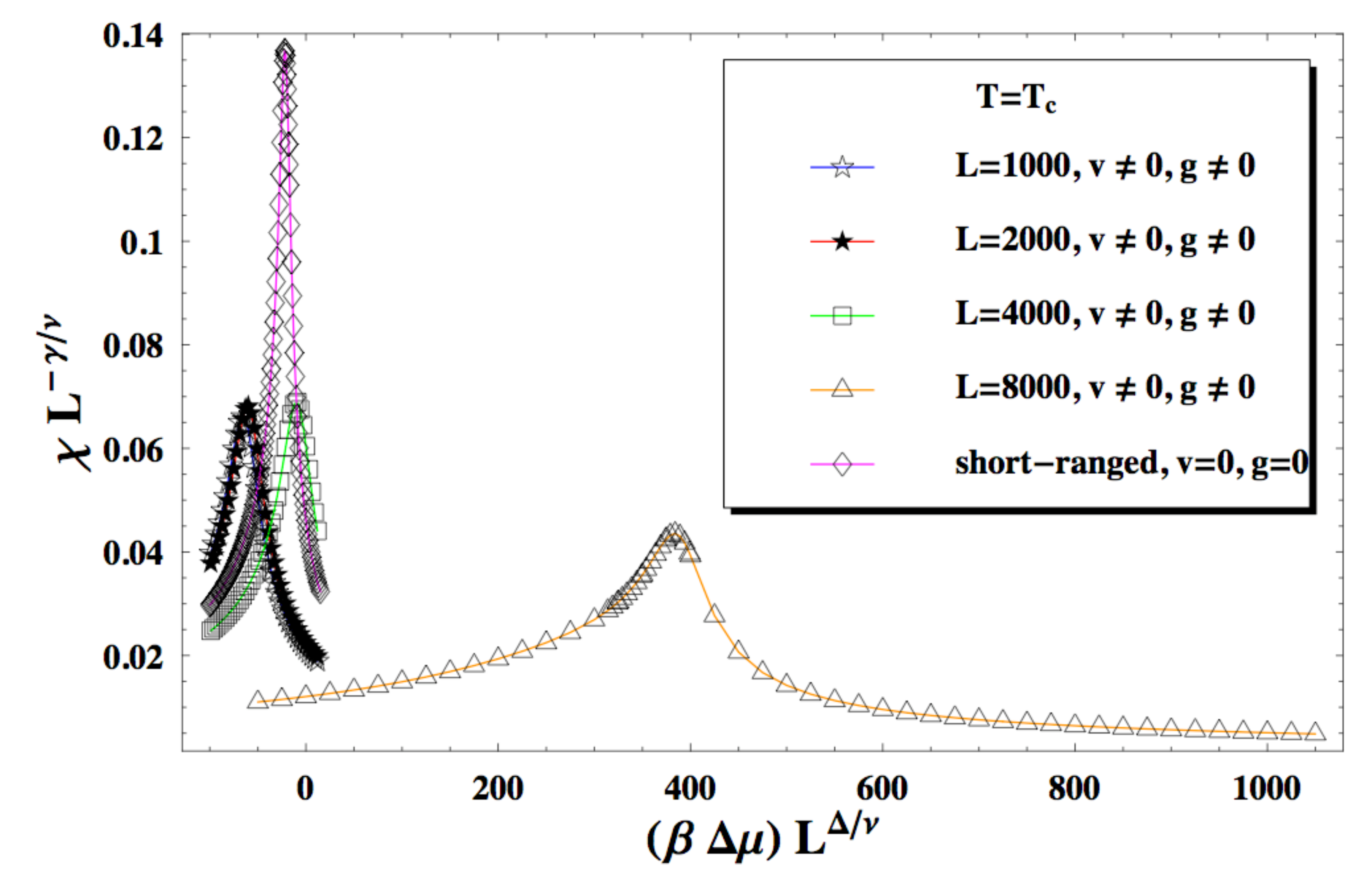}
\caption{Color online. Finite-size behavior of the  susceptibility for a
film with $L=1000, 2000, 4000$ and $8000$ layers as a function of the
scaling variable $x_\mu=\Delta\mu L^{\Delta/\nu}$ in the case in which both
the intrinsic van der Waals interactions as well as the presence of gravity are taken into account. For a comparison the behavior of a system with
completely short-ranged interaction is also presented.
  \label{chi_verus_rho_all_disc}}
\end{figure}

 Figure \ref{chi_verus_rho_all_disc} summarizes the behavior of the  susceptibility for such
films as exemplified with films with thicknesses $L=1000, 2000, 4000$ and $8000$ layers. The susceptibility is shown for temperature equal to that the bulk critical temperature, $T=T_c$, plotted as a function of  the scaling variable $x_\mu=\Delta\mu L^{\Delta/\nu}$. The quantity $\Delta \mu$ is the excess chemical potential where $\Delta \mu >0$ stabilizes the liquid phase of the fluid.  Both the intrinsic van der Waals pair interactions between the molecules of the fluid and the van der Waals interaction between the fluid and the constituents of the two substrate plates are taken into account. For comparison, the behavior of the finite-size susceptibility of a system with completely short-ranged interactions is also presented. We conclude that the behavior of the susceptibility in realistic nonpolar fluid systems, in which both van der Waals interactions and gravity are present, will always differ from the corresponding behavior of short-ranged systems, which constitute the standard theoretical model for such systems. As might have been expected, for realistic films with moderate film thickness $L$, the van der Waals interactions lead to noticeable  differences in the behavior of the susceptibility from what one finds for a short-ranged system. For large $L$, gravity gives rise to the the dominant effect. For the system studied here, consisting of He films bounded by Au surfaces it turns out that ``moderate'' thickness is a width of up to $L=4000$ layers; the smaller the thickness the stronger the effect due to van der Waals interactions. It turns out that $L=8000$ represents a ``thick'' film, for which the principal effects giving rise to the deviation from the short-ranged behavior of the susceptibility are due to the presence  of a gravitational field in the region occupied by the fluid. At the bulk critical point $T=T_c$ and $\Delta \mu=0$ for $L=8000$ layers gravity gives rise to  the appearance of a very large gas-like region that
includes the middle of the film which splits the two liquid-like regions near the solid substrate planes that bounds the film; see Fig. \ref{density}. This gas-like region occupies the greater part of both the upper and the lower half of the fluid system, but is larger near the upper bounding surface. This leads to the result that at $T=T_c$ and $\Delta \mu=0$  and for $L$ large enough, the average density of the finite system is less than the critical density of the bulk system. Thus, in order to achieve coexistence between the fluid and the gas-like states of the system one needs to apply a positive excess chemical potential. This in turn leads to a new finite-size coexistence line in thin films; see the topmost line in Fig. \ref{phasediagram}, which is different from the ones obtained on the basis of studies of fluid systems with no gravity present \cite{NF82,NF83,BLM2003,BL91}  where such a shift of the coexistence line is always in the direction of negative $\Delta \mu$.

Finally, we mention that in order to verify our lattice model approach we have in the case of a fully short-ranged system derived analytically and compared with numerical calculations on a lattice model the behavior of the local and total susceptibilities. We find excellent  agreement between the two models. The details  are given in Appendix \ref{app:chianalytic}.

\acknowledgments
A portion of this work was carried out at the Jet Propulsion Laboratory under a contract with the National Aeronautics and Space Administration.

DD acknowledges partial financial support under grant TK171/08
of the Bulgarian NSF. JR acknowledges partial support from the National Science Foundation.

\appendix

\section{Estimation of the gravity constant in terms of the parameters of the model}
\label{gravity}

As we see from Eqs. (\ref{hz})  and (\ref{gcpomegafilmfinal}), gravity introduces a field-like contribution into the grand canonical potential normalized per area. This contribution is reflected by the term
\begin{equation}\label{G}
\beta H_G \equiv \frac{1}{2}\beta g\sum_{z=0}^L  z \phi(z)= \frac{1}{2}\frac{K}{K_c}\left[ \beta_c g \sum_{z=0}^L  z \phi(z) \right].
\end{equation}
Here we measure $z$ from the bottom of the fluid layer, which is positioned perpendicularly to the gravitational force. We work in units in which $|\phi(z)|<1$ is a dimensionless number, with $\phi(z)=1$ corresponding to one atom of $^3$He or $^4$He occupying a unit cell with volume $a_0^3$, where $a_0$ is the average distance between the helium atoms at the critical point. One can think of $a_0$ as being the lattice spacing of the lattice model. In \cite{DRB2007} we estimated $a_0=4.9$ {\AA} for $^3$He and $a_0=4.2$ {\AA} for $^4$He; the basic material specific characteristics of the two isotopes of the helium needed for the current estimations are summarized in Table \ref{table}.  Thus, the actual physical density, corresponding to $\phi(z)=1$ is equal to  $\rho=n u/a_0^3$, where $n$ is equal to $3$ for $^3$He and $4$ for $^4$He, and $u$ is the atomic mass unit; $1u=1.6605 \times 10^{-27}$ kg. Next,  with $g=9.81 \,m/s^2=9.81 \,{\rm J}/{(\rm kg \times m)}$ and with $z=l_z \, a_0 \, 10^{-10}\,{\rm m}$, where $l_z$ is an integer that denotes the layer occupied by the corresponding particle is, one obtains
\begin{eqnarray}\label{gHe3}
\beta_c \rho g z &=& \left(2.2 \times
10^{22}{\rm J}^{-1}\right)  \times \left(3 \times 1.66054 \times
10^{-27}\frac{{\rm kg}}{a_0^3}\right)\nonumber \\
&& \times \left(9.81  \,\frac{{\rm
J}}{{\rm kg \times m}}\right) \times \left(l_z \times \, 4.9
\,\times 10^{-10}\,{\rm m}\right) \nonumber \\
&=&0.527 \times 10^{-12}
\frac{l_z}{a_0^3}
\end{eqnarray}
for $^3$He, and
\begin{eqnarray}\label{gHe4}
\beta_c \rho g z &=&  \left(1.4
\times 10^{22}{\rm J}^{-1}\right)  \times \left(4 \times 1.66054
\times 10^{-27}\frac{{\rm kg}}{a_0^3}\right)\nonumber \\
&& \times \left(9.81
\,\frac{{\rm J}}{{\rm kg \times m}}\right) \times \left(l_z \times
\, 4.2 \,\times 10^{-10}\,{\rm m}\right) \nonumber \\
&=& 0.383 \times 10^{-12}
\frac{l_z}{a_0^3}
\end{eqnarray}
for $^4$He.
\begin{table}[h]
\caption{Material-specific characteristics for $^3$He and $^4$He used to estimate parameters of the model investigated in the article.}
\label{table}
\begin{center}
\begin{tabular}{|c|c|c|c|c|c|}
  \hline \hline
          & $a_0$     & $T_c$ &$\beta_c$ & $\rho$ & $\rho_c$\\
  \hline \hline
  $^3$He &  $a_0=4.9$ {\AA} & 3.3 {\rm K}    & $2.2 \times 10^{22} {\rm J^{-1}}$    & $3u/a_0^3$   & 0.041 \rm{g/cm$^3$} \\
  $^4$He &  $a_0=4.2$ {\AA} & 5.2 {\rm K}    & $1.4 \times 10^{22}{\rm  J^{-1}}$    & $4u/a_0^3$  & 0.069 {\rm g/cm$^3$} \\
  \hline
\end{tabular}
\end{center}
\end{table}
In the above estimates we have taken into account the fact that \cite{DRB2007} $$\beta_c=2.2 \times 10^{22}\,{\rm J}^{-1}$$ for $^3$He and $$\beta_c=1.4 \times 10^{22}\, {\rm J}^{-1}$$ for $^4$He  (see Table \ref{table}). We thus conclude that in our units one has to take the gravitational constant (times $\beta_c$) to be
$$(\beta_c g) \equiv g_{3}=0.527 \times 10^{-12} \quad \text{for} \quad ^3{\rm He},$$ and
$$(\beta_c g) \equiv g_{4}=0.383 \times 10^{-12} \quad \text{for} ^4{\rm He.}$$ Note, however, that in the equation for the order parameter profile (\ref{mpfilm}), as well as in the excess grand potential (\ref{gcpomegafilmfinal}), which are written in terms of $\phi$ one always has  a factor of $1/2$ in front of gravitational constant. This should not be forgotten when performing a numerical evaluation of the gravity effect.

Taking into account the fact that the critical part of the free energy  near $T_c$ behaves as $t^{2-\alpha}$, with $\alpha =0$ within the mean-field approximation, it is clear that gravitational effects can be felt in the thermodynamic behavior of the system when $t^2\sim 10^{-12}$.  This implies that one must explore relative temperature deviations of the order of  $t\sim 10^{-6}$ in order to be able to observe severe gravitational effects in the finite-size behavior of thermodynamic quantities. Obviously, the larger $L$, the stronger will be those effects. However, in such a case the observation of the finite-size properties of the studied quantities will be more challenging since they emerge when $tL^{1/\nu}=O(1)$. Thus, one needs to find a proper balance in the size of the system in order to observe both finite-size and gravity effects. In \cite{BHLD2007} the conclusion has been made that we that
gravity has a much more pronounced effect near liquid-gas
critical points than for the $^4$He lambda transition. One can also compare the strength of the expected gravity effects in different substances near their respective critical points -- see, e.g. \cite{S75} and Table I in \cite{BHLD2007}. There such a comparison is performed for $^3$He, SF$_6$,  Xe and CO$_2$ in terms of the so-called ``gravity scale height in a fluid"  $H_0$, with $H_0\equiv P_c/(\rho_c g)$, where $P_c$ is the critical pressure of the fluid \cite{S75}. The smaller $H_0$ the larger the gravity effect. It turns out that $H_0$ is smallest for $^3$He, i.e. the gravity effect there is the strongest. $H_0$ steadily increases for the sequence  SF$_6$,  Xe and CO$_2$, i.e. the strength of the gravity effects diminishes in these substances in the sequence they are ordered in the current text. Our own estimates support this observation; we predict that the gravity effects in the critical behavior of $^3$He will be stronger than in $^4$He, since the corresponding effective parameter reflecting the strength of the gravity in the model $g_3$ is larger than $g_4$; see above.

\section{Response of the system with short range interactions---Ginzburg Landau approach} \label{app:chianalytic}

In the case in which all the interactions in the system are short-ranged and in the absence of gravity, Eq. (\ref{eqcont}) for the order parameter profile in a continuum system can be written in the standard form
\begin{equation} \label{eqcontsr}
-\frac{d^2 \phi}{dz^2}+\hat{a} \phi + u \phi^3=\hat{h},
\end{equation}
where
\begin{equation}\label{srconst}
\hat{a}\equiv \frac{1}{c_2^{nn}K}\left(1-\frac{K}{K_c}\right),
\quad u\equiv\frac{1}{3 c_2^{nn}K}, \quad \mbox{and} \quad \hat{h}\equiv\frac{h}{c_2^{nn}K}.
\end{equation}
With the substitution
\begin{equation}\label{substm}
\phi(z)=\sqrt{\frac{2}{u}}\; m(z)
\end{equation}
the above equation becomes
\begin{equation} \label{eqcontsrm}
-\frac{d^2 m}{dz^2}+\hat{a} m + 2 m^3=\bar{h},
\end{equation}
where
\begin{equation}\label{hbar}
\bar{h}=\sqrt{\frac{u}{2}}\; \hat{h}.
\end{equation}
The equations for the layer response function
\begin{equation}\label{localsusdef}
\chi_l(z)\equiv \frac{\partial \phi(z)}{\partial \hat{h}}{{\Bigg |}_{\hat{h}=0}}=\frac{\partial m(z)}{\partial \bar{h}}{{\Bigg |}_{\bar{h}=0}}
\end{equation}
then reads
\begin{equation} \label{eqcontsrchi}
-\frac{d^2 \chi_l}{dz^2}+(\hat{a}+6m^2) \chi_l =1.
\end{equation}
Introducing the scaling variables
\begin{equation}\label{scavar}
x_t= \hat{a} L^2, \qquad x_h=\bar{h} L^3, \qquad \zeta=z/L,
\end{equation}
one can rewrite the equations (\ref{eqcontsrm}) and (\ref{eqcontsrchi}) for the order parameter $m(z)$ and of the local susceptibility $\chi_l(z)$ into equations for the scaling functions
\begin{equation}\label{scafunm}
X_m(z|x_t,x_h)=L\; m(z|x_t,x_h)
\end{equation}
and
\begin{equation}\label{scafunchi}
X_\chi(z|x_t,x_h)=L^{-2}\; \chi_l(z|x_t,x_h)
\end{equation}
 of these quantities. One obtains
\begin{equation} \label{eqcontsrXm}
-\frac{d^2 X_m}{d\zeta^2}+x_t X_m + 2 X_m^3=x_h,
\end{equation}
and
\begin{equation} \label{eqcontsrXchi}
-\frac{d^2 X_\chi}{d \zeta^2}+(x_t+6 X_m^2) X_\chi =1,
\end{equation}
respectively.
Because boundary conditions are identical at both surfaces of the
film system, the solutions of the above equations have to satisfy  $\phi'(L/2)=0$ and
$\chi'(L/2)=0$ or, equivalently, $X_m'(\zeta=1/2)=0$ and $X_\chi'(\zeta=1/2)=0$, i.e., the middle of the system will be an inflection point for all physical quantities which depend on the distance from the boundaries. Note that despite the fact that there is no explicit dependence of Eq. (\ref{eqcontsrXchi}) on $x_h$, the scaling function, $X_\chi$, does depend on $x_h$ because $X_m$, which is a solution of Eq. (\ref{eqcontsrXm}), depends on that parameter and $X_m$ enters into Eq. (\ref{eqcontsrXchi}).

In the remainder of this appendix we will be interested in the behavior of $X_\chi$ on $z$ and $x_t$ for $x_h=0$. Then, the solution $X_m(z|x_t)$ of Eq. (\ref{eqcontsrXm}) is known. From Eqs. (\ref{mz}) and (\ref{mzd}) one has
\begin{enumerate}
\item[a)] when $x_t\ge -\pi^2$
\begin{equation}\label{Xmpp}
X_m(\zeta|x_t)= 2 K(k)\frac{{\rm dn}[2 K(k)\zeta;k]}{{\rm sn}[2 K(k)\zeta;k]},
\end{equation}
where $0\leq \zeta\leq 1$, and $k\in \mathbb{R}$ is to be determined form Eq. (\ref{tk}).

\item[b)] when $x_t \le -\pi^2$
\begin{equation}\label{Xmpn}
X_m(\zeta|x_t) = \frac{2 K(\bar{k})}{{\rm sn}[2 K(\bar{k})\zeta;\bar{k}]},
\end{equation}
where $0\leq \zeta\leq 1$, and $\bar{k}\in \mathbb{R}$ is to be determined form Eq. (\ref{tkb}).
\end{enumerate}
\begin{figure}[htbp]
\begin{center}
\includegraphics[width=3in]{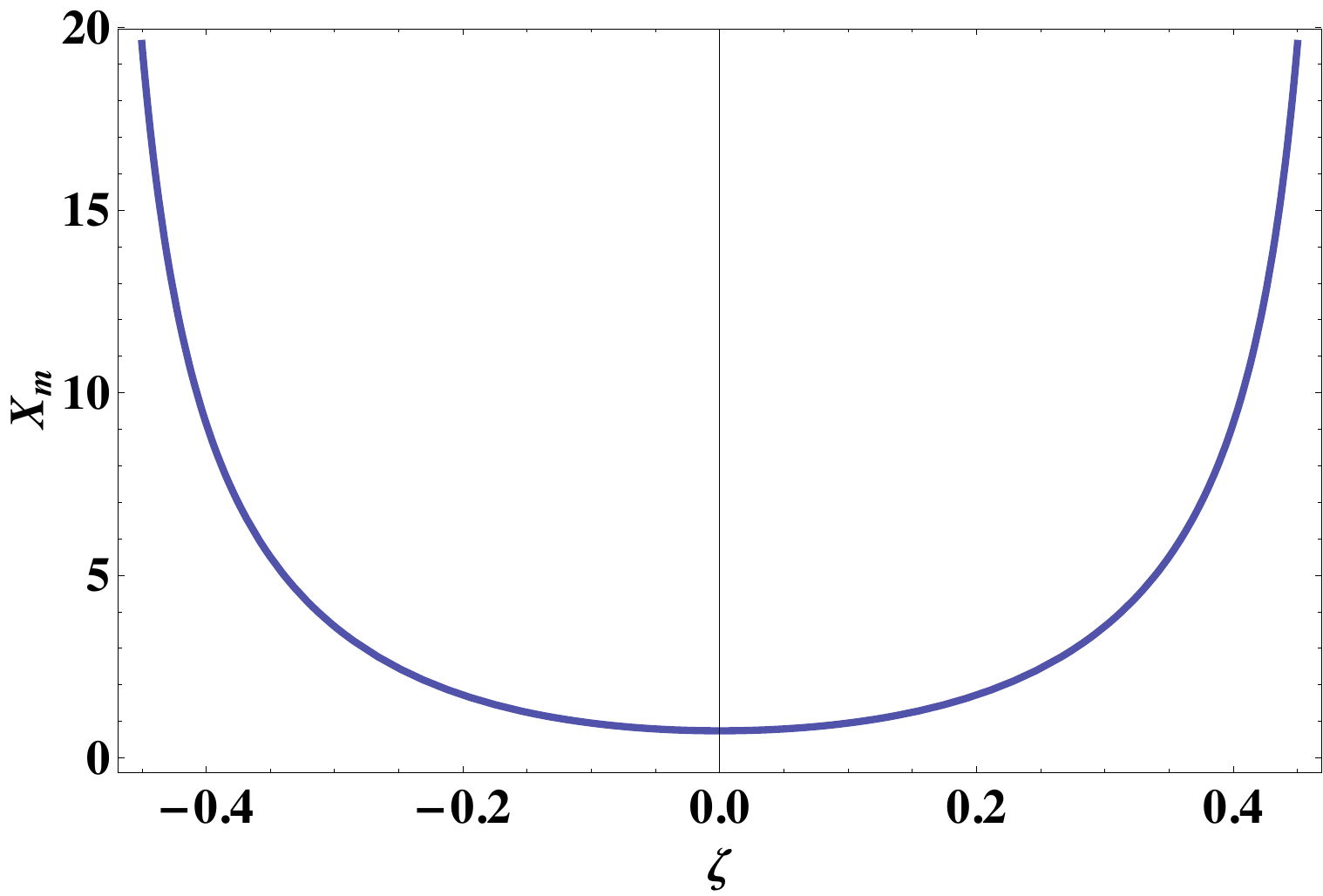}
\caption{Color online. Plot of the profile function $X_m(\zeta|x_t)$ - see Eq. (\ref{Xmmiddle}).}
\label{fig:Xm}
\end{center}
\end{figure}

In order to utilize the symmetry of the problem it is helpful to move the coordinate frame so that the origin of the system is at the midpoint of the film. Taking into account that \cite{GR73}
\begin{equation}\label{elfunperprop}
\frac{{\rm dn}[u+K(k);k]}{{\rm sn}[u+K(k);k]}=\frac{k'}{{\rm cn}(u;k)},
\end{equation}
and that
\begin{equation}\label{elfuncnprop}
{\rm cn}(iu;k')=1/{\rm cn}(u;k),
\end{equation}
we obtain
\begin{equation}\label{Xmmiddle}
X_m(\zeta|x_t)=X_{m,0} \;{\rm cn}[i 2K(k) \zeta;k']
\end{equation}
where $\zeta\in [-1/2,1/2]$ and
\begin{equation}\label{Xm0def}
X_{m,0} \equiv 2 k' K(k)=2 K(\bar{k}).
\end{equation}
Since ${\rm cn}(0;k)=1$, one has
\begin{equation}\label{Xm0}
X_{m,0}=X_m(0|x_t).
\end{equation}
A typical behavior of $X_m(\zeta|x_t)$ is shown on Fig. \ref{fig:Xm}.

Given the scaling function $X_m$ all that one has to do to determine $X_\chi(\zeta|x_t)$ is to solve Eq. (\ref{eqcontsrXchi}) with the boundary conditions $X_\chi(\pm 1/2)=0$ and $dX_\chi(\zeta)/d\zeta=0$ for $\zeta=0$.

According to the general theory of differential equations of second order
\begin{figure}[htbp]
\begin{center}
\includegraphics[width=3in]{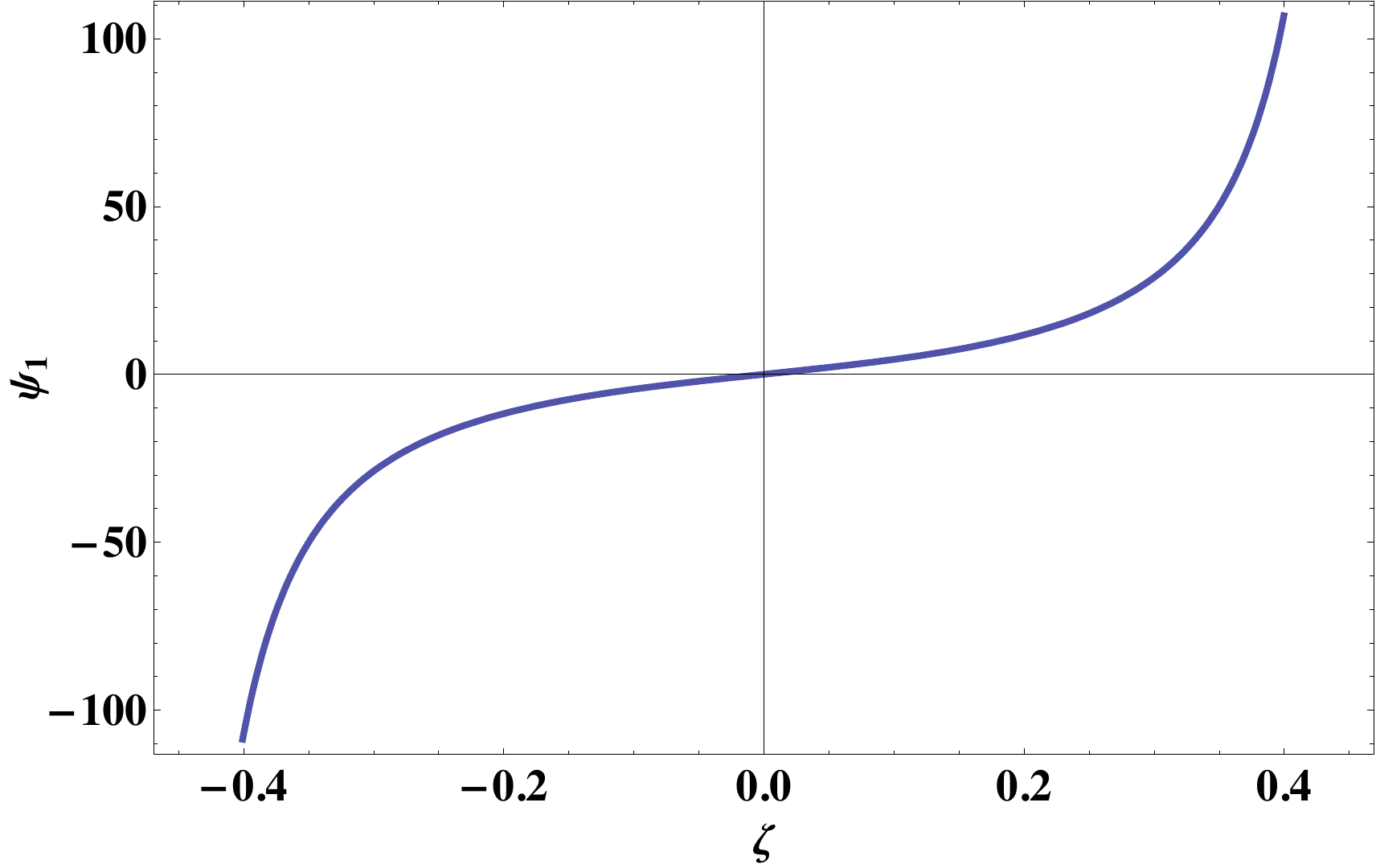}
\caption{Color online. Plot of the function $\psi_1(\zeta|x_t)$ - see Eq. (\ref{firstsolexpl}).}
\label{fig:psi1}
\end{center}
\end{figure}
\begin{equation}\label{Xchigen}
X_\chi(\zeta|x_t)=c_1 \psi_1 (\zeta|x_t)+ c_2 \psi_2 (\zeta|x_t)+ c_i \psi_i (\zeta|x_t),
\end{equation}
where $c_1$, $c_2$ and $c_i$ are constants, $\psi_1$ and $\psi_2$ are linearly independent solutions of the homogeneous equation
\begin{equation} \label{eqcontsrXchihom}
-\frac{d^2 X_\chi}{d \zeta^2}+(x_t+6 X_m^2) X_\chi =0,
\end{equation}
and $\psi_i$ is a particular solution of the inhomogeneous Eq. (\ref{eqcontsrXchi}). It is easy to check that
\begin{equation}\label{firstsol}
\psi_1(\zeta|x_t)=\frac{d}{d\zeta}X_m(\zeta|x_t)\equiv \dot{X}_m(\zeta|x_t)
\end{equation}
is a solution of Eq. (\ref{eqcontsrXchihom}). Explicitly, one has
\begin{eqnarray}\label{firstsolexpl}
\lefteqn{\psi_1(\zeta|x_t)= -i \frac{X_{m,0}^2}{k'}  {\rm sn}\left[i 2K(k) \zeta;k'\right] {\rm dn}\left[i 2K(k) \zeta;k'\right]} \nonumber \\
&=&  -i \frac{X_{m,0}^2}{k'}  {\rm sn}\left[i  \frac{X_{m,0}}{k'} \zeta;k'\right] {\rm dn}\left[i  \frac{X_{m,0}}{k'} \zeta;k'\right].
\end{eqnarray}
A typical behavior of the solution of the homogeneous equation $\psi_1(\zeta|x_t)$ is shown on Fig. \ref{fig:psi1}.
Because of the symmetry of the problem,
\begin{equation}\label{symXchi}
X_\chi(-\zeta|x_t)=X_\chi(\zeta|x_t).
\end{equation}
Since, see Eq. (\ref{firstsolexpl}),
\begin{equation}\label{sympsi1}
\psi_1(-\zeta|x_t)=-\psi_1(\zeta|x_t),
\end{equation}
one concludes that $c_1=0$; see Eq. (\ref{Xchigen}).
\begin{figure}[htbp]
\begin{center}
\includegraphics[width=3in]{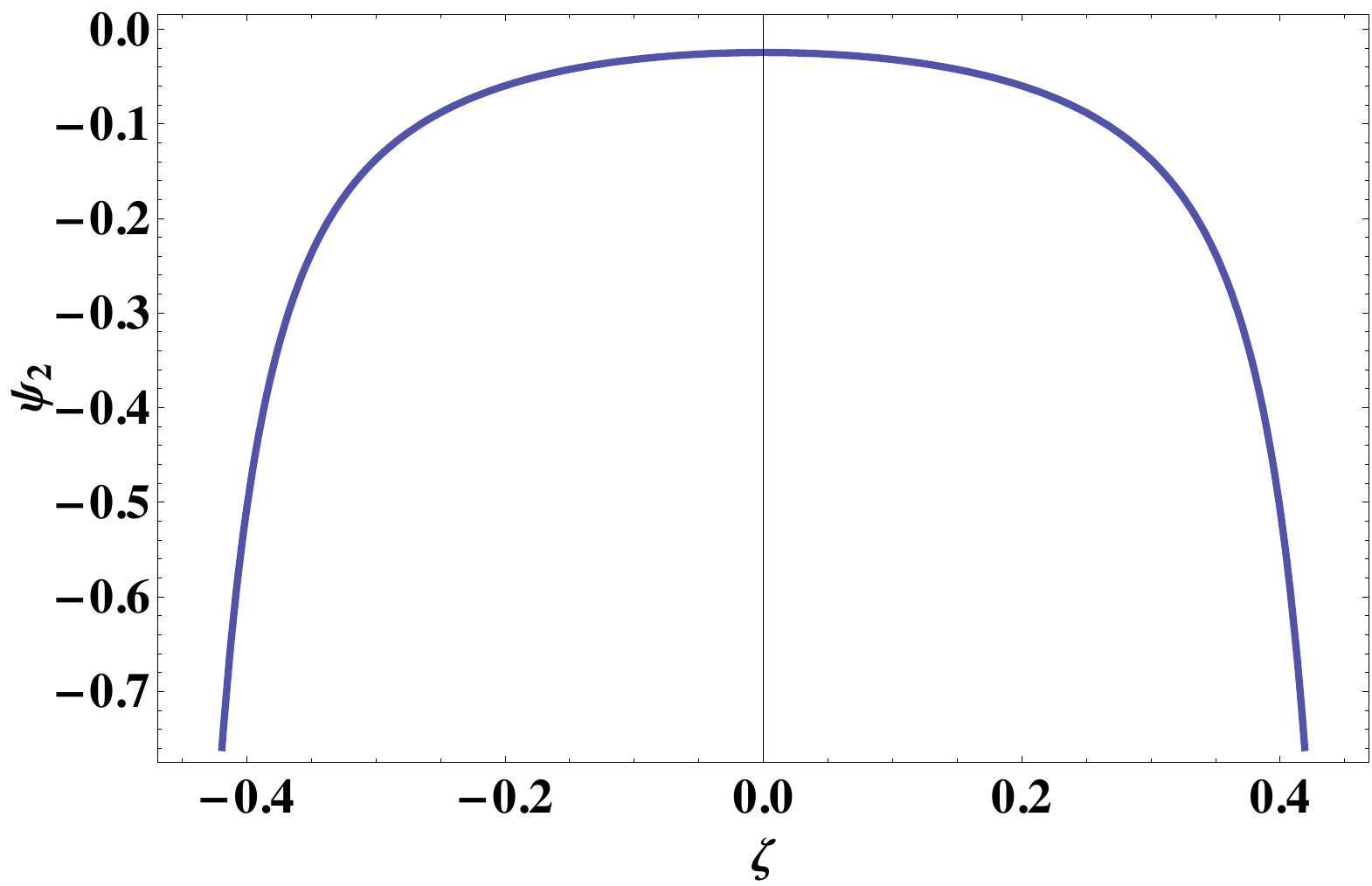}
\caption{Color online. Plot of the function $\psi_2(\zeta|x_t)$ - see Eq. (\ref{secondsolexpl}).}
\label{fig:psi2}
\end{center}
\end{figure}

Following Abel \cite{abel,boyce}, we construct $\psi_2(\zeta|x_t)$ via
\begin{equation}\label{secondsol}
\psi_2(\zeta|x_t)=\psi_1(\zeta|x_t)\int \frac{d\zeta}{[\psi_1(\zeta|x_t)]^2}.
\end{equation}
Obviously, $\psi_2(\zeta|x_t)$ is an even function with respect to $\zeta$. Again, because of the symmetry of this system, one has $d\psi_2(\zeta|x_t)/d\zeta =0$ at $\zeta=0$.
Next, it is easy to check that the Wronskian of the solutions of $\psi_1$ and $\psi_2$
\begin{equation}
W \equiv \psi_1(\zeta) \frac{d \psi_2(\zeta)}{d\zeta} - \psi_2(\zeta) \frac{d \psi_1(\zeta)}{d\zeta}
\label{eq:Wval}
\end{equation}
is equal to one, i.e., the two solutions $\psi_1$ and $\psi_2$ are linearly independent.
Explicitly, performing the integration for $\psi_2$, one obtains
\begin{eqnarray}
\lefteqn{\psi_2(\zeta|x_t) = -\frac{k'}{k^2 X_{m,0}^3}\left\{ {\rm dn}\left(i \frac{X_{m,0}}{k'}\,\zeta;k' \right) {\rm sn}\left(i \frac{X_{m,0}}{k'}\,\zeta;k' \right) \right.} \nonumber \\
&& \left[k'(1-2k'^2){\rm E}\left({\rm am}\left(i \frac{X_{m,0}}{k'}\,\zeta;k' \right);k'\right)-ik^2X_{m,0}\zeta\right] \nonumber  \\
&&+k'{\rm cn}\left(i \frac{X_{m,0}}{k'}\,\zeta;k' \right)\\
\times && \left.\left[k'^2+(1-2k'^2)\;{\rm dn}\left(i \frac{X_{m,0}}{k'}\,\zeta;k' \right)2\right]\right\}. \nonumber
\label{secondsolexpl}
\end{eqnarray}
A typical behavior of the solution of the homogeneous equation $\psi_2(\zeta|x_t)$ is shown on Fig. \ref{fig:psi2}.
\begin{figure}[htbp]
\begin{center}
\includegraphics[width=3in]{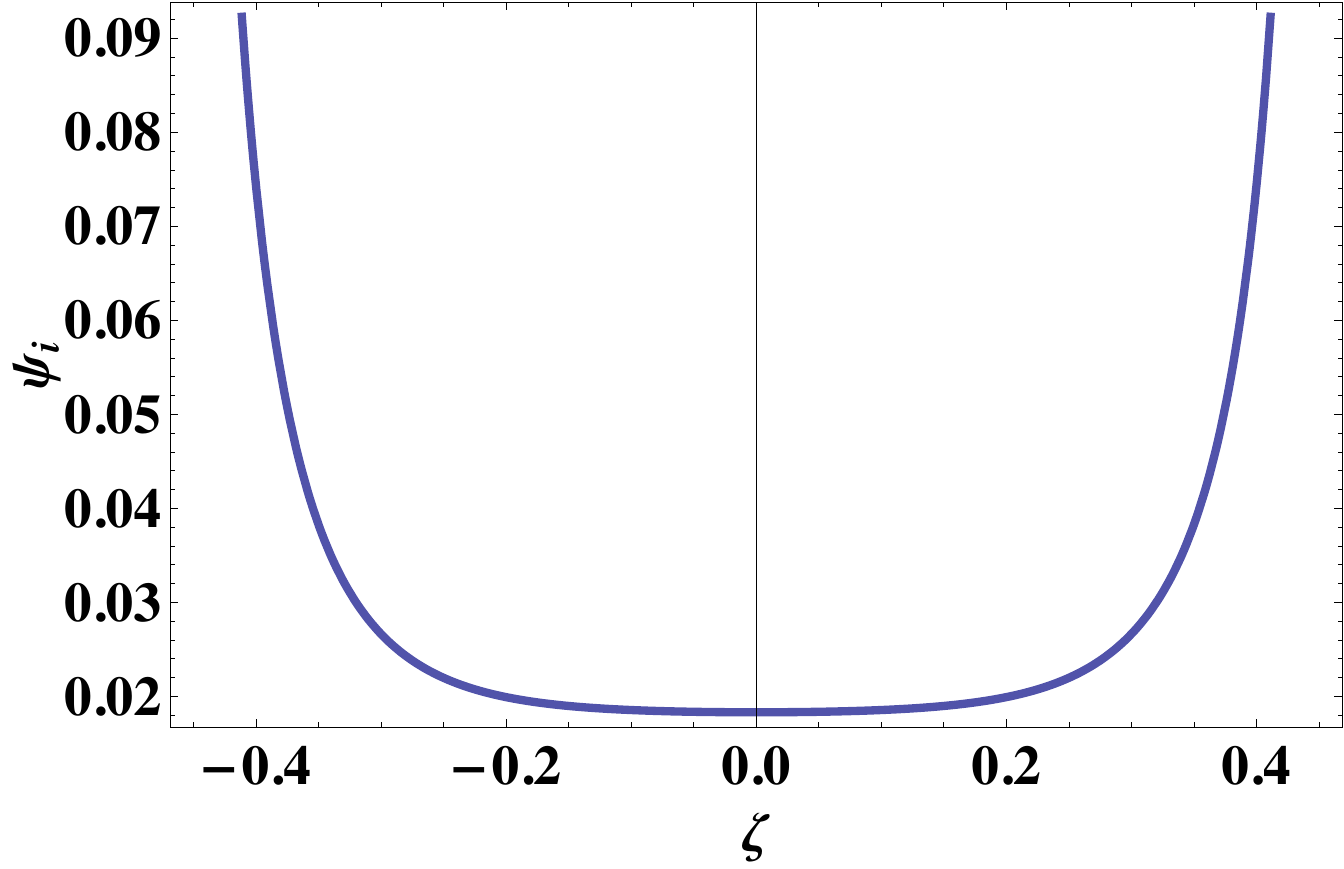}
\caption{Color online. Plot of the function $\psi_i(\zeta|x_t)$ - see Eq. (\ref{partsolexpl}).}
\label{fig:psii}
\end{center}
\end{figure}

Again following Abel \cite{abel,boyce}, one finds the particular solution of the inhomogeneous equation
expressed in terms of $\psi_1$ and $\psi_2$ via
\begin{equation}\label{partsol}
\psi_i(\zeta|x_t)=\psi_1(\zeta|x_t)\int \psi_2(\zeta|x_t)d\zeta-\psi_2(\zeta|x_t)\int \psi_1(\zeta|x_t)d\zeta.
\end{equation}
Taking into account Eq. (\ref{firstsol}), one can show that
\begin{equation}\label{partsolfinal}
\psi_i(\zeta|x_t)=-\psi_1(\zeta|x_t)\int \frac{X_m(\zeta|x_t)}{[\psi_1(\zeta|x_t)]^2}d\zeta.
\end{equation}
Using this result and that fact that $\psi_1$ is a solution of the homogeneous equation (\ref{eqcontsrXchihom}), one can show that $\psi_i$ is indeed a solution of Eq. (\ref{eqcontsrXchi}). Furthermore, since the constant on the right-hand side of Eq. (\ref{eqcontsrXchi}) is equal to one, we have $c_i=1$. One can explicitly determine $\psi_1(\zeta|x_t)$. Performing the integration in Eq. (\ref{partsolfinal}), one obtains
\begin{equation}\label{partsolexpl}
\psi_i(\zeta|x_t)=-\frac{k'^{\,2}}{X_{m,0}^2}\left\{1-2\,{\rm dn}\left[i \frac{X_{m,0}}{k'}\,\zeta;k'\right]^2\right\},
\end{equation}
with $d\psi_i(\zeta|x_t)/d\zeta =0$ at $\zeta=0$.
A typical behavior of the particular solution of the inhomogeneous equation $\psi_i(\zeta|x_t)$ is shown on Fig. \ref{fig:psii}.
\begin{figure}[htbp]
\begin{center}
\includegraphics[width=3in]{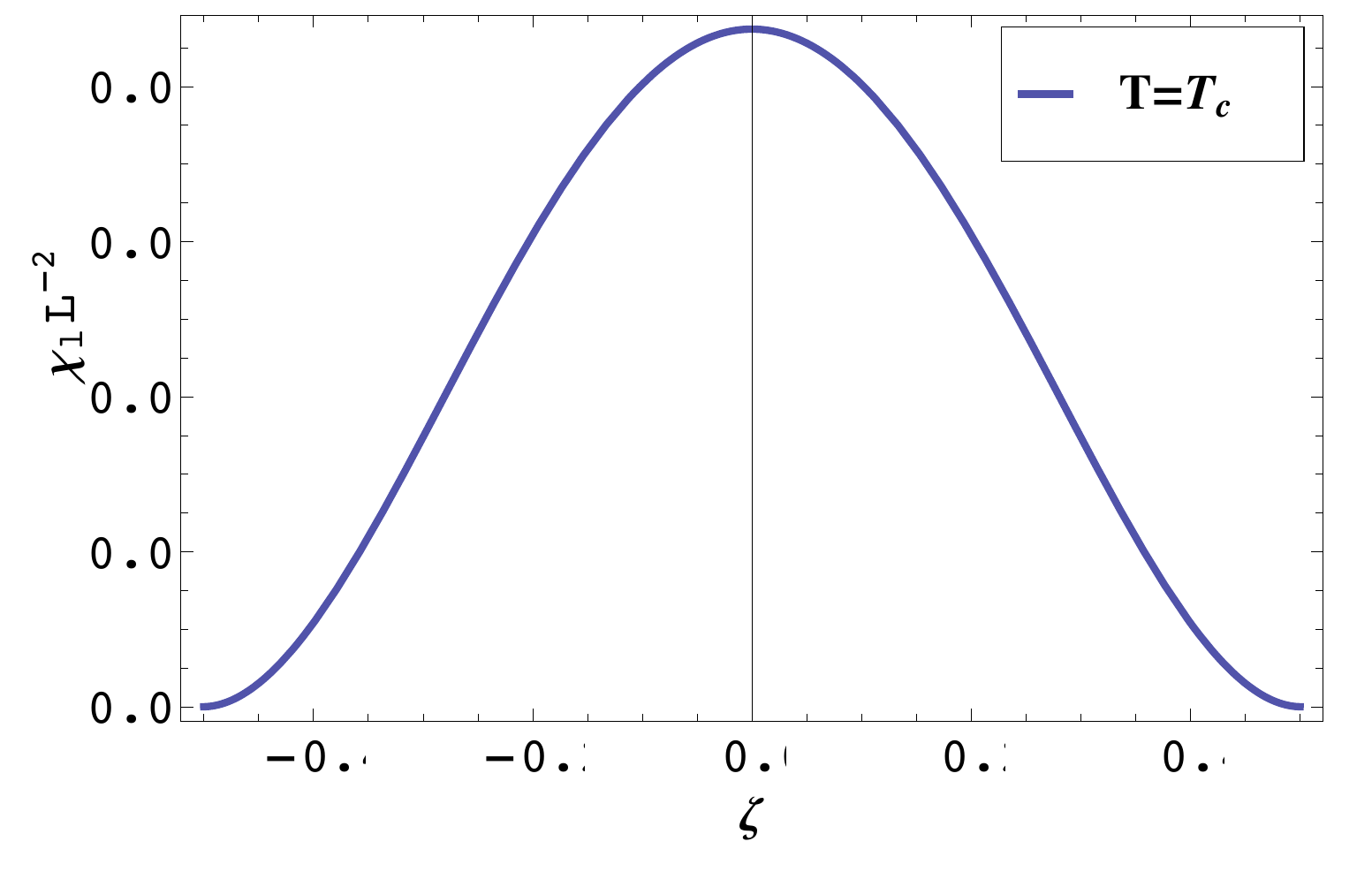}
\caption{Color online. Plot of the function $\chi_l(\zeta)$.}
\label{fig:chilatsr}
\end{center}
\end{figure}
Since $c_i=1$ and $c_1=0$ from $X_\chi(1/2|x_t=0)$ we determine that $c_2(x_t)=-\lim_{\zeta\to 1/2} [\psi_i(\zeta|x_t)/\psi_2(\zeta|x_t)]$. Explicitly, after series of manipulations one can show that
\begin{equation}\label{c2expl}
c_2(x_t)=\frac{4 k'^2 k^2 K\left(k\right)}{k'^2
   K\left(k\right)+\left(k^2-k'^2\right) E\left(k\right)}.
\end{equation}

The behavior of the scaling function $X_\chi$ at $T=T_c$ obtained from the continuum approach described above is shown on Fig. \ref{fig:XchiTc} in the main text.
One can also calculate the local susceptibility profile from the lattice model approach. The result for $L=1000$ is shown on Fig. \ref{fig:chilatsr}.
One can easily check that the curves in both figures completely overlap each other if $L^{-2}\chi_l$, calculated via the lattice model, is multiplied by $c_r=0.212$. The last is very close to what one theoretically predicts from the mapping of the lattice model onto the continuum one. Inspection shows that  $c_r=(3c_2^{\rm nn}K_c)^{-1}=0.198$. The above demonstrates that the lattice model correctly reproduces the universal scaling function in the case in which that function can be calculated exactly and, thus, is trustworthy for predicting properties of the local and total susceptibility in cases when no analytical results are available. Such an instance, considered in the main text, is one of a system in which both gravity and van der Waals interactions are present.

In addition to the scaling function of the local susceptibility one can also derive in an analytic closed form the scaling function $X$ of the total susceptibility. Starting from
\begin{equation}\label{relXlX}
X(x_t)=2\int_0^{1/2}X_\chi(\zeta|x_t)d\zeta,
\end{equation}
one derives
\begin{equation}\label{Xfinal}
X(x_t)=\frac{c2(x_t)/K\left(k\right)+K\left(k\right)-2 E\left(k\right)}{4
   K^3\left(k\right)}.
\end{equation}
The behavior of this function is illustrated in Fig. \ref{fig:totsuscont} in the main text.

In order to compare with the corresponding result from the lattice calculations one must take into account the fact that, due to the difference in definitions of the field variable, $X$ must be rescaled  by $1/c_r$. Additionally, taking into account the fact that $(1-K/K_c)L^{1/\nu}=c_r x_t$ (see Eqs. (\ref{srconst}) and (\ref{scavar})) one obtains the result shown in Fig. \ref{fig:totsuscontlat} in the main text. We are thus able to conclude that  the lattice model produces a result in a perfect agreement with the analytic solution in Eq. (\ref{Xfinal}) for the behavior of the total susceptibility.

\end{document}